\documentclass[a4paper,aps,11pt,nofootinbib,showpacs]{revtex4}

\usepackage{amsmath, amssymb,amsfonts}
\usepackage[dvips]{graphicx}
\usepackage[dvips]{color} 

\begin{document}
\title{\Large{\bf Stationary Closed Strings in Five-dimensional Flat Spacetime}}

\hfill{OCU-PHYS 368}

\hfill{AP-GR-99}


\author{$^{1,2}$Takahisa Igata} 
\email{igata@sci.osaka-cu.ac.jp}
\author{$^{2}$Hideki Ishihara}
\email{ishihara@sci.osaka-cu.ac.jp}
\author{$^{2}$Keisuke Nishiwaki}
\email{nisiwaki@sci.osaka-cu.ac.jp}

\affiliation{
$^{1}$Department of Physics, Kinki University, Osaka 577-8502, Japan\\
$^{2}$Department of Mathematics and Physics, Graduate School of Science, Osaka City University, Osaka 558-8585, Japan}

\begin{abstract}
We investigate stationary rotating closed Nambu-Goto strings in five-dimensional flat spacetime. The stationary string is defined as a worldsheet that is tangent to a timelike Killing vector. Nambu-Goto equation of motion for the stationary string is reduces to the geodesic equation on the orbit space of the isometry group action generated by the Killing vector. We take a linear combination of a time-translation vector and space-rotation vectors as the Killing vector, and explicitly construct general solutions of stationary rotating closed strings in five-dimensional flat spacetime. We show a variety of their configurations and properties.
\end{abstract}

\maketitle

\section{Introduction}
Existence of extended objects such as strings attracts much attention in the connection of unified theories and cosmologies. Microscopic strings are one of fundamental objects describing elementary particles including gravity in superstring theories. On the other hand, many unified models of particles predict formation of macroscopic strings by spontaneous symmetry breaking in the early universe, called cosmic strings~\cite{Vilenkin-Shellard:1994}. In both fields of modern physics it is important issue to discuss dynamics of strings.

Recently, attention is focused on a possibility that strings are formed at the end of brane inflation~\cite{Dvali:1998pa, Alexander:2001ks, Burgess:2001fx, Dvali:2001fw} in the context of braneworld scenarios~\cite{ArkaniHamed:1998rs, ArkaniHamed:1998nn, Randall:1999ee, Randall:1999vf} in which our universe might have large extra dimensions. Such strings may have astronomical sizes, and are called cosmic superstrings~\cite{Jones:2002cv, Sarangi:2002yt} (and see Ref.~\cite{Copeland:2009ga} for review).

Properties of strings in higher-dimensional spacetime are quite different in the case of four-dimensional spacetimes. One of crucial features of the string in higher-dimensional spacetime is reduction of reconnection probability~\cite{Polchinski:1988cn, Jackson:2004zg} (and see also Ref.~\cite{Polchinski:2004ia}), while the probability is almost unity in four-dimensional spacetime. One of the qualitative understanding is that if a string apparently intersects in our visible universe then strings can still avoid intersection in extra dimensions~\cite{Dvali:2003zj}. Another crucial feature appears in cusp formation, which is a point where string segment reaches the speed of light and its extrinsic curvature diverges. Cusps are formed on closed string loops generically in four-dimensional spacetimes~\cite{Kibble:1982cb, Turok:1984cn}. In contrast, the probability of cusp formation is to be measure zero in higher-dimensional spacetimes.

Let us review the discussion of cusp formation for closed Nambu-Goto strings. We consider a string that is described by the two-dimensional worldsheet $\Sigma$ embedded in $D$-dimensional target spacetime $(M, g_{\mu\nu})$. The worldsheet $\Sigma$ is represented by the mapping functions $y^\mu(\tau, \sigma)$, where $y^\mu$ are coordinates of $M$, and $\tau$ and $\sigma$ are parameters on $\Sigma$. We assume that dynamics of a string is governed by the Nambu-Goto action
\begin{align}
S_{\rm NG}=-\mu\int_{\Sigma}d^2\sigma \sqrt{-\gamma},
\end{align}
where $\mu$ is the string tension and $\gamma$ is the determinant of the induced metric on $\Sigma$, which is given by
\begin{align}
\gamma_{ab}=g_{\mu\nu}\frac{\partial y^{\mu}}{\partial \sigma^{a}}\frac{\partial y^{\nu}}{\partial \sigma^{b}},
\end{align}
where $\sigma^0=\tau$ and $\sigma^1=\sigma$.

Let us consider a solution of the closed Nambu-Goto string in the $D$-dimensional flat spacetime with the Cartesian coordinates $(t, y^1, y^2, \ldots, y^{D-1})$. We use the conformal gauge such that
\begin{align}
\gamma_{01}=0, \quad \gamma_{00}+\gamma_{11}=0.
\label{eq:conformalgauge}
\end{align}
Then the equations of motion are simply
\begin{align}
(-\partial_\tau^2+\partial_\sigma^2) y^\mu=0.
\label{eq:wave_eq}
\end{align} 
We can fix the residual gauge freedom by setting $t=\tau$. In this case, a string $\mbox{\boldmath $y$}=(y^1,\cdots y^{D-1})$ in the flat spacetime is described by right moving and left moving waves as
\begin{align}
\mbox{\boldmath $y$}(\tau, \sigma)=\frac12\left[\mbox{\boldmath $F$}(u_+)+\mbox{\boldmath $G$}(u_-)\right],
\end{align}
where $\mbox{\boldmath $F$}$ and $\mbox{\boldmath $G$}$ are arbitrary functions of $u_{\pm}:=\tau\pm\sigma$. The gauge conditions require $\mbox{\boldmath $F$}$ and $\mbox{\boldmath $G$}$ should satisfy
\begin{align}
\mbox{\boldmath $F$}'(u_+)^2=\mbox{\boldmath $G$}'(u_-)^2=1,
\label{eq:unitsphere}
\end{align}
where the prime denotes the differentiation with respect to the argument. That is, $\mbox{\boldmath $F$}'$ and $\mbox{\boldmath $G$}'$ are confined to the $(D-2)$-dimensional unit sphere, known as the ``Kibble-Turok sphere".

For closed strings we require the periodicity $\mbox{\boldmath $y$}(\tau, \sigma)=\mbox{\boldmath $y$}(\tau, \sigma+\sigma_p)$ leads to
\begin{align}
\mbox{\boldmath $F$}'(u_++\sigma_p)=\mbox{\boldmath $F$}'(u_+), \quad
\mbox{\boldmath $G$}'(u_--\sigma_p)=\mbox{\boldmath $G$}'(u_-).
\end{align}
 
Furthermore, in the center of mass frame of the string, where the total momentum vanishes, these vector functions should satisfy
\begin{align}
\int\mbox{\boldmath $F$}'d\sigma=\int\mbox{\boldmath $G$}'d\sigma =0.
\end{align}
Hence, $\mbox{\boldmath $F$}'$ and $\mbox{\boldmath $G$}'$ draw closed curves on the Kibble-Turok sphere, and their averaged positions coincide with the center of the sphere.

In the case $D=4$, such two closed curves on the two-dimensional sphere intersect quite generally. The intersecting points correspond to the emergence of cusps~\cite{Kibble:1982cb, Turok:1984cn} (see also \cite{Vilenkin-Shellard:1994}). It is proposed that cosmic string cusps can give rise to gravitational wave emission that is detectable by future experiments~\cite{Berezinsky:2000vn, Damour:2000wa, Damour:2001bk}. However, in the higher-dimensional spacetime where $D>4$, intersecting probability of the curves on the $(D-2)$-dimensional sphere is measure zero. Therefore, we conclude that cusp formation is not generic event in the higher-dimensional spacetime. The absence of cusp formation in higher-dimensions make difference in gravitational wave emission compared with the four-dimensional case. Detectability of burst gravitational radiation from cosmic superstring near cusps is discussed in Refs.~\cite{Damour:2004kw, O'Callaghan:2010ww, O'Callaghan:2010sy, O'Callaghan:2010hq}.

One of the simplest and fundamental problem in string dynamics is finding stationary configurations in a background spacetime. Stationary open string configuration has been studied extensively, after the pioneering work by Burden and Tassie~\cite{Burden:1982zb, Burden:1984xk}. Non-trivial configurations of a stationary string in four-dimensional flat spacetime were discovered in various approaches~\cite{deVega:1996mv, Frolov:1996xw, Ogawa:2008qn, Burden:2008zz}. Indeed, in the flat spacetimes, general solutions to Eq.~\eqref{eq:wave_eq} are easily obtained, but it is non-trivial problem to select out stationary solutions that satisfy the non-linear conditions~\eqref{eq:conformalgauge}.

The stationary strings, defined in Sec.~\ref{sec:2}, are in one of the classes of cohomogeneity-one strings as was discussed in Ref.~\cite{Ishihara:2005nu}. All possible cohomogeneity-one strings in four-dimensional flat spacetime were completely classified into seven families~\cite{Ishihara:2005nu}. It was shown that equations for all types of cohomogeneity-one strings are exactly solvable~\cite{Kozaki:2009jj}. Such classification was also done in five-dimensional anti-de Sitter background~\cite{Koike:2008fs}.

In four-dimensional flat spacetime, we have no stationary closed string solution to the Nambu-Goto equations because of the cusp formation. Indeed, in four-dimensions, closed string solutions are constructed explicitly in flat background~\cite{Vilenkin:1981kz}, de Sitter background~\cite{Basu:1991ig, Basu:1992ue, DeVega:1992xc}, cosmological background~\cite{Akdeniz:1989ti, Li:1993qc}, and black hole background~\cite{Lousto:1993vj, deVega:1993hq}, but all of them are not stationary. Even if not cusp formation, string loops collapse to a point or a double line in a period because string tension works inward. However, in higher-dimensions, stationary closed loops are not forbidden. The recent investigations of string dynamics in higher-dimensional spacetimes have revealed that stationary loop solutions do exist. This is closely related with the absence of cusp formation. Special solutions of stationary loop configuration were provided, and their properties were discussed in five-di
 mensional flat spacetime~\cite{BlancoPillado:2007iz, Igata:2009dr}. Furthermore, separability of equations for stationary string configurations, and the solutions were demonstrated in higher-dimensional black hole spacetime~\cite{Igata:2009dr, Igata:2009fd}. The purpose of the present paper is to study stationary configurations of closed strings extensively in higher-dimensional flat spacetime.

The organization of this paper is as follows. In the following section, we introduce a stationary string in five-dimensional flat spacetime by use of a timelike Killing vector, and reduce Nambu-Goto action to a geodesic action in four-dimensional Riemannian space determined by the Killing vector. Using the Hamilton-Jacobi method, we demonstrate separation of variables of the geodesic equations to ordinary differential equations of single variables due to existence of a rank-2 Killing tensor. In Sec.~\ref{sec:3}, we show stationary configuration of closed strings by solving the equations. This section consists of two parts: special cases and general cases. The special cases include solutions of stationary toroidal spiral strings, which are obtained in the previous works~\cite{BlancoPillado:2007iz, Igata:2009dr, Igata:2009fd}. We also consider a special class of stationary strings called planar strings that lie on two-dimensional plane. In general cases, we demonstrate typical configurations of closed strings. The final section presents summary and discussion. We use the sign convention $-++++$ for the metric, and units in which $c=G=1$.

\section{Stationary Strings in Five-dimensional Flat Spacetime}
\label{sec:2}
Let $(M, g_{\mu\nu})$ be a spacetime that possesses a Killing vector $\xi^\mu$, where  $\mu, \nu$ run from $0$ to $4$. Then we consider a string with a worldsheet $\Sigma$ tangent to $\xi^\mu$, which is called a cohomogeneity-one string~\cite{Ishihara:2005nu}. In particular, if $\xi^\mu$ is a timelike Killing vector, the cohomogeneity-one string associated with $\xi^\mu$ is called a stationary string.

Let $G$ be the isometry group generated by $\xi^\mu$. The set of all integral curves of 
$\xi^\mu$ defines the orbit space of $G$ denoted by $M/G$. The metric on $M/G$ is naturally introduced by
\begin{align}
h_{\mu\nu}=g_{\mu\nu}-\xi_{\mu}\xi_{\nu}/f,
\label{eq:h}
\end{align}
where $f=\xi^\mu \xi_\mu$. As discussed in Ref.~\cite{Frolov:1988zn}, the problem of finding cohomogeneity-one Nambu-Goto string reduces to solving the geodesic equation in the orbit space $M/G$ with the norm weighted metric $fh_{\mu\nu}$ derived by the action
\begin{align}
S=\int\sqrt{-fh_{\mu\nu}dy^\mu dy^\nu}.
\label{eq:action}
\end{align}
In what follows, we analyze this system in the five-dimensional flat background.
We introduce coordinates $\bar{t}$, $\bar{\rho}$, $\bar{\phi}$, $\bar{\zeta}$, $\bar{\psi}$ in the five-dimensional flat spacetime so that the metric takes the form
\begin{align}
ds^2=-d\bar{t}^2+d\bar{\rho}^2+\bar{\rho}^2d\bar{\phi}^2+d\bar{\zeta}^2+\bar{\zeta}^2d\bar{\psi}^2.
\end{align}
We consider a linear combination of a time translation Killing vector and two rotation Killing vectors in the five-dimensional flat spacetime in the form
\begin{align}
\xi^\mu=\partial^\mu_{\bar{t}}+\alpha \partial^\mu_{\bar{\phi}}+\beta\partial^\mu_{\bar{\psi}},
\label{eq:xi}
\end{align}
where $\alpha$ and $\beta$ are constants. Then we have
\begin{align}
f=-1+\alpha^2\bar{\rho}^2+\beta^2\bar{\zeta}^2.
\label{eq:f}
\end{align}
We can assume that $\alpha$ and $\beta$ are non-negative because negative case is obtained by the reflection of coordinates $\bar{\phi}$ and $\bar{\psi}$. Hence, by Eqs.~\eqref{eq:h}, \eqref{eq:xi} and \eqref{eq:f}, the metric, $-fh_{\mu\nu}$, is expressed in the form
\begin{align}
ds^2_{M/G}
&=-f h_{\mu\nu}dy^\mu dy^\nu \\
&=-f\left[d\bar{\rho}^2 + \bar{\rho}^2(d\bar{\phi}-\alpha d\bar{t})^2+d\bar{\zeta}^2 + \bar{\zeta}^2(d\bar{\psi}-\beta d\bar{t})^2\right]+\left[\alpha \bar{\rho}^2(d\bar{\phi}-\alpha d\bar{t})+\beta \bar{\zeta}^2 (d\bar{\psi}-\beta d\bar{t})\right]^2.
\nonumber
\end{align}
By the new coordinates $x^i$ given by
\begin{align}
(x^1, x^2, x^3, x^4)
=(\rho, \phi, \zeta, \psi)
=(\bar{\rho}, \bar{\phi}-\alpha\bar{t}, \bar{\zeta}, \bar{\psi}-\beta\bar{t}),
\end{align}
the four-dimensional Riemannian metric components covering $M/G$ are
\begin{align}
ds^2_{M/G}
=-f h_{ij}dx^{i}dx^{j}
=-f\left[d\rho^2+\rho^2 d\phi^2+d\zeta^2+\zeta^2d\psi^2\right]+(\alpha \rho^2d\phi+\beta \zeta^2 d\psi)^2.
\label{eq:metric}
\end{align}

Note that $-fh_{ij}$ is singular at points satisfying $f=0$, where the scalar curvature,
\begin{align}
R=\frac{6[\alpha^2+\beta^2-\alpha^2\beta^2(\rho^2+\zeta^2)]}{(-1+\alpha^2\rho^2+\beta^2\zeta^2)^3},
\end{align}
diverges.

We solve the geodesic equations in $(M/G, -fh_{ij})$ by the Hamilton-Jacobi method. It is convenient to rewrite the action \eqref{eq:action}, which has reparameterization invariance, to the equivalent action,
\begin{align}
S=\frac12\int d\sigma \left(-N^{-1}fh_{ij}x'^ix'^j+\kappa N\right),
\end{align}
where the prime denotes the differentiation with respect to the parameter $\sigma$, and $\kappa$ is a constant, and an arbitrary function $N$ of $\sigma$ is a Lagrange multiplier that is related to the reparameterization invariance of geodesic orbits. The Hamiltonian of the system is of the form 
\begin{align}
H&=-\frac{N}{2}(f^{-1}h^{ij}p_ip_j+\kappa)\cr
&=-\frac{N}{2f} \left(p_\rho^2+p_\zeta^2+\frac{1-\alpha^2\rho^2}{\rho^2}L_1^2+\frac{1-\beta^2\zeta^2}{\zeta^2}L_2^2-2\alpha\beta L_1 L_2+\kappa f\right),
\label{eq:H}
\end{align}
where the canonical momentum $p_i$ conjugate to $x^i$ is defined by $p_i=-N^{-1}fh_{ij}x'^j$, and $L_1=p_{\phi}$ and $L_2=p_{\psi}$ denote constants of motion.

We attempt to apply the Hamilton-Jacobi method to Eq.~\eqref{eq:H}. Let $S$ be Hamilton's principal function, which is a function of the parameter $\sigma$ and coordinates  $x^i$, and the Hamilton-Jacobi equation is given by
\begin{align}
\frac{\partial S}{\partial \sigma}+\frac{N}{2} \frac{h^{ij}}{f}\frac{\partial S}{\partial x^i} \frac{\partial S}{\partial x^j}=0.
\label{eq:H-J}
\end{align}
We suppose that $S$ is a complete solution, which includes the same number of constants as the dimensions of $M/G$, i.e., four constants in this case, and takes the form of the complete separation of variables, 
\begin{align}
S=\frac{\kappa}{2}\chi+L_1 \phi+L_2 \psi+S_\rho(\rho)+S_\zeta(\zeta),
\label{eq:S}
\end{align}
where $\chi$ is a function of $\sigma$ such that $\chi'=N$, and $S_{\rho}$ and $S_{\zeta}$ are functions that depend only on $\rho$ and $\zeta$, respectively. Substitution of this expression into Eq.~\eqref{eq:H-J} yields
\begin{align}
-&\left(\frac{d S_{\zeta}}{d \zeta}\right)^2
-\frac{1-\beta^2\zeta^2}{\zeta^2} L_2^2
+\alpha \beta L_1L_2
+\frac{1-2\beta^2\zeta^2}{2}\kappa\cr
=&\left(\frac{d S_{\rho}}{d \rho}\right)^2
+\frac{1-\alpha^2\rho^2}{\rho^2} L_1^2
-\alpha \beta L_1L_2
-\frac{1-2\alpha^2\rho^2}{2}\kappa = K,
\label{eq:K}
\end{align}
where $K$ is a separation constant. The existence of the quadratic constant $K$ in $p_i$ implies that the metric \eqref{eq:metric} admits the rank-2 Killing tensor $K^{ij}$ of the form
\begin{align}
K^{ij}=\frac{1}{2f}
\bigg[&(1-2\beta^2\zeta^2)\partial_\rho^i\partial_\rho^j
-(1-2\alpha^2\rho^2)\partial_\zeta^i\partial_\zeta^j
-2\alpha\beta(\alpha^2\rho^2-\beta^2\zeta^2)\partial_\phi^{(i}\partial_\psi^{j)}\cr&
+\frac{(1-\alpha^2\rho^2)(1-2\beta^2\zeta^2)}{\rho^2}\partial_\phi^i\partial_\phi^j-\frac{(1-\beta^2\zeta^2)(1-2\alpha^2\rho^2)}{\zeta^2}\partial_\psi^i\partial_\psi^j\bigg].
\end{align}
From Eqs.~\eqref{eq:S} and \eqref{eq:K}, we obtain the complete solution by virtue of the separability of the Hamilton-Jacobi equation,
\begin{align}
S=\frac{\kappa}{2}\chi+L_1\phi+L_2\psi+s_\rho \int d\rho \sqrt{\Theta_\rho}+s_\zeta \int d\zeta \sqrt{\Theta_\zeta},
\end{align}
where
\begin{align}
\Theta_\rho=-\frac{1-\alpha^2\rho^2}{\rho^2}L_1^2+\alpha \beta L_1L_2+\frac{1-2\alpha^2\rho^2}{2}\kappa+K,\\
\Theta_\zeta=-\frac{1-\beta^2\zeta^2}{\zeta^2} L_2^2+\alpha \beta L_1L_2+\frac{1-2\beta^2\zeta^2}{2}\kappa-K,
\end{align}
where $s_{\rho}$ and $s_{\zeta}$ denote $\pm1$. By setting partial derivatives of $S$ with respect to $\kappa$, $L_1$, $L_2$, $K$ to zero, we obtain
\begin{align}
&\chi
=\frac{s_{\rho}}{2}\int d\rho \frac{1-2\alpha^2\rho^2}{\sqrt{\Theta_{\rho}}}
+\frac{s_{\zeta}}{2}\int d\zeta \frac{1-2\beta^2\zeta^2}{\sqrt{\Theta_{\zeta}}}, \\
&\phi
=\frac{s_{\rho}}{2}\int d\rho \frac{2(\rho^{-2}-\alpha^2)L_1-\alpha \beta L_2}{\sqrt{\Theta_{\rho}}}
-\frac{s_{\zeta}}{2}\int d\zeta \frac{\alpha\beta L_2}{\sqrt{\Theta_{\zeta}}}, \\
&\psi
=\frac{s_{\zeta}}{2}\int d\zeta \frac{2(\zeta^{-2}-\beta^2)L_2-\alpha \beta L_1
}{\sqrt{\Theta_{\zeta}}}
-\frac{s_{\rho}}{2}\int d\rho \frac{\alpha\beta L_1}{\sqrt{\Theta_{\rho}}}, \\
&s_{\rho} \int \frac{d\rho}{\sqrt{\Theta_{\rho}}}
=s_{\zeta} \int \frac{d\zeta}{\sqrt{\Theta_{\zeta}}}.
\end{align}
It is useful for analysis to express the first-order differential equations,
\begin{align}
\rho'^2&=\frac{N^2}{f^2}\left(- \frac{1-\alpha^2\rho^2}{\rho^2} L_1^2
+\frac{1-2\alpha^2\rho^2}{2}\kappa
+K+\alpha \beta L_1L_2\right),
\label{eq:preEOM_rho}\\
\zeta'^2&=\frac{N^2}{f^2}\left(-\frac{1-\beta^2\zeta^2}{\zeta^2} L_2^2
+\frac{1-2\beta^2\zeta^2}{2}\kappa
-K+\alpha \beta L_1L_2\right),
\label{eq:preEOM_zeta}\\
\phi'&=-\frac{N}{f}\left(\frac{1-\alpha^2\rho^2}{\rho^2}L_1-\alpha\beta L_2\right),
\label{eq:preEOM_phi}\\
\psi'&=-\frac{N}{f}\left(\frac{1-\beta^2\zeta^2}{\zeta^2}L_2-\alpha\beta L_1\right).
\label{eq:preEOM_psi}
\end{align}

\section{Solutions for Stationary Closed Strings}
\label{sec:3}
In this section, we present stationary string solutions explicitly by integrating the equations \eqref{eq:preEOM_rho}--\eqref{eq:preEOM_psi}. In particular, we focus on stationary closed strings and discuss their configuration and properties.

Suppose that the Lagrange multiplier $N$ takes the form\footnote{Note that the gauge choice \eqref{eq:gauge_condition} is not the conformal gauge.}
\begin{align}
N=-f=1-\alpha^2\rho^2-\beta^2\zeta^2,
\label{eq:gauge_condition}
\end{align}
then Eqs.~\eqref{eq:preEOM_rho}--\eqref{eq:preEOM_psi} are simplified as
\begin{align}
&\rho'^2+U = P,
\label{eq:rho_eq}\\
&\zeta'^2+ V = Q,
\label{eq:zeta_eq}\\
&\phi'=\frac{L_1}{\rho^2}-\alpha C,
\label{eq:phi_eq}\\
&\psi'=\frac{L_2}{\zeta^2}-\beta C,
\label{eq:psi_eq}
\end{align}
where $C$ is defined by
\begin{align}
C=\alpha L_1+\beta L_2,
\label{eq:C}
\end{align}
and $P$ and $Q$ are constants given by
\begin{align}
P=\alpha L_1 C+\frac{1}{2}+K,
\label{eq:P}\\
Q=\beta L_2 C+\frac{1}{2}-K.
\label{eq:Q}
\end{align}
The functions $U$ and $V$ in Eqs.~\eqref{eq:rho_eq} and \eqref{eq:zeta_eq} are defined by
\begin{align}
U&=\frac{L_1^2}{\rho^2}+\alpha^2\rho^2,
\label{eq:U}\\
V&=\frac{L_2^2}{\zeta^2}+\beta^2\zeta^2,
\label{eq:V}
\end{align}
respectively.

Equations \eqref{eq:rho_eq} and \eqref{eq:zeta_eq} show that configuration of a stationary string is described by a couple of one-dimensional particle motions 
in the effective potentials $U$ and $V$ with \lq energies\rq\ $P$ and $Q$. In the case $\alpha=0$ (or $\beta=0$), the second term of $U$ (or $V$) vanishes, and then the stationary strings are unbounded and have infinite lengths. On the other hand, for non-vanishing $\alpha$ and $\beta$, stationary strings are confined in the range
\begin{align}
0\leq \rho_-\leq \rho(\sigma)\leq \rho_+, \\
0\leq \zeta_- \leq \zeta(\sigma)\leq \zeta_+,
\end{align}
where 
\begin{align}
\rho^2_\pm&=\frac{P\pm \sqrt{P^2-4 l_1^2}}{2\alpha^2},
\label{eq:rho_pm}\\
\zeta^2_\pm&=\frac{Q\pm\sqrt{Q^2-4 l_2^2}}{2\beta^2},
\label{eq:zeta_pm}
\end{align}
and $l_1=\alpha L_1, l_2=\beta L_2$.

For real non-negative $\rho_\pm$ and $\zeta_\pm$, we have $P\geq 0, P^2-4l_1^2\geq 0, Q\geq 0$, and $Q^2-4 l_2^2\geq 0$. From these inequalities we obtain
\begin{align}
&-\frac{1+C^2}{2} \leq l_1, l_2 \leq \frac{1+C^2}{2},
\label{eq:parameter_range_l}\\
&-\frac{1+C^2}{2} \leq l_2-l_1 \leq \frac{1+C^2}{2},
\label{parameter_range_l_2-l_1}\\
&-\frac{1+|C|(1+C^2)}{2} \leq K \leq \frac{1+|C|(1+C^2)}{2}.
\label{parameter_range_K}
\end{align}

Furthermore, from the requirement for stationarity of a string, the Killing vector $\xi^\mu$ must be timelike, i.e.,
\begin{align}
f=-1+\alpha^2\rho^2(\sigma)+\beta^2\zeta^2(\sigma)<0,
\label{eq:timelike_condition}
\end{align}
then, we see
\begin{align}
\alpha^2\rho_+^2+\beta^2\zeta_-^2<1, \quad \mbox{and}\quad
\alpha^2\rho_-^2+\beta^2\zeta_+^2<1.
\end{align}
Substituting Eqs.~\eqref{eq:rho_pm} and \eqref{eq:zeta_pm} into these inequality we obtain
\begin{align}
C^2<1-\Delta, \quad \mbox{and}\quad C^2<1+\Delta,
\end{align}
where
\begin{align}
\Delta=\sqrt{P^2-4 l_1^2}-\sqrt{Q^2-4 l_2^2}.
\end{align}
Therefore, we have
\begin{align}
-1<C=l_1+l_2<1.
\label{eq:C_range}
\end{align}

From the inequalities~\eqref{eq:parameter_range_l}--\eqref{parameter_range_K}, and \eqref{eq:C_range}, we have
\begin{align}
&-1<l_1, l_2<1,\\
&-1<l_2-l_1<1, \\
&-\frac{3}{2}<K<\frac{3}{2}.
\end{align}
A stationary solution in this class is closed if its geodesic in the orbit space is closed.

\subsection{Special Cases}
\subsubsection{Toroidal Spiral Strings}
In this section, we discuss special solutions of Eqs.~\eqref{eq:rho_eq}--\eqref{eq:psi_eq} that are given by constant $\rho$ and $\zeta$. Such strings are called the stationary toroidal spiral strings~\cite{Igata:2009fd, Igata:2009dr} (see also Ref.~\cite{BlancoPillado:2007iz}).

Let us discuss stationary toroidal spiral string solutions. The minima of each effective potential \eqref{eq:U} and \eqref{eq:V}, which are given by $dU/d\rho=0$ and $dV/d\zeta=0$, are
\begin{align}
\rho_0^2=\frac{\left|l_1\right|}{\alpha^2}, \\
\zeta_0^2=\frac{\left|l_2\right|}{\beta^2}.
\end{align}
We assume that $l_1\neq 0$ and $l_2 \neq 0$, i.e., $\rho_0\neq 0$ and $\zeta_0 \neq 0$.

Since $\rho'=\zeta'=0$, we see that
\begin{align}
P+Q=C^2+1=2(|l_1|+|l_2|).
\label{eq:P+Q}
\end{align}
In the case $l_1l_2<0$, from Eq.~\eqref{eq:P+Q} and the definition of $C$ in Eq.~\eqref{eq:C} we obtain
\begin{align}
l_1&= \frac{(C+1)^2}{4},
\label{eq:l_1}\\
l_2&=-\frac{(C-1)^2}{4},
\label{eq:l_2}
\end{align}
where we have assumed $l_1>0$ without loss of generality because discussion of $l_1<0$ is equivalent to replacing $\sigma$ by $-\sigma$. From $P-Q$, we see easily that
\begin{align}
K=\frac{C(3-C^2)}{4}.
\end{align}
On the other hand, if $l_1 l_2>0$, we find $C=\pm1$, which contradicts with Eq.~\eqref{eq:C_range}. Then, we consider the case $l_1>0$, $l_2<0$ in this subsection.

The equations for the angular variables, $\phi'=\alpha\left(1-C\right)$, $\psi'=-\beta\left(1+C\right)$, can be immediately integrated, and then
\begin{align}
&\rho_0 = \frac{1+C}{2\alpha},\\
&\zeta_0 = \frac{1-C}{2\beta},\\
&\phi(\sigma) = \alpha \left(1-C\right)\sigma,\\
&\psi(\sigma) = -\beta \left(1+C\right)\sigma,
\end{align}
where the integration constants are suitably chosen. After reparameterizations $(\alpha/(1+C)$, $\beta/(1-C), (1-C^2) \sigma)\to(\alpha,\beta,\sigma)$ the solutions are simply represented as
\begin{align}
&\rho_0=\frac{1}{2\alpha},
\label{eq:rho0}\\
&\zeta_0=\frac{1}{2\beta},
\label{eq:zeta0}\\
&\phi(\sigma)=\alpha\sigma,
\label{eq:phiTSS}\\
&\psi(\sigma)=-\beta\sigma.
\label{eq:psiTSS}
\end{align}
The string described by the solution has a shape of toroidal spiral that lies on the two-dimensional torus with the metric
\begin{align}
ds^2_{\mathrm{S}^1\times \mathrm{S}^1}=\frac{d\phi^2}{4\alpha^2}+\frac{d\psi^2}{4\beta^2}.
\end{align}
The toroidal spiral string solutions have two parameters $\alpha$ and $\beta$.\footnote{The two parameters $\alpha$ and $\beta$ are related to the parameters $\alpha$ and $L$ in Ref.~\cite{Igata:2009dr} by $\alpha \leftrightarrow \beta/\alpha$ and $L \leftrightarrow 1/(4\alpha \beta)$.}

Both angular solutions must be periodically identified with periods $\sigma_1=2\pi/\alpha$ and $\sigma_2=2\pi/\beta$, i.e., we have
\begin{align}
\phi(\sigma+n_1 \sigma_1)&=\phi(\sigma), \\
\psi(\sigma+n_2 \sigma_2)&=\psi(\sigma),
\end{align}
where $n_1$ and $n_2$ are relatively prime integers that denote winding numbers of the string in $\rho$-$\phi$ and $\zeta$-$\psi$ plane, respectively.

When we restrict attention to closed loops, we must require $n_1\sigma_1=n_2\sigma_2$, then we have
\begin{align}
\frac{n_2}{n_1}=\frac{\sigma_1}{\sigma_2}=\frac{\beta}{\alpha}=\frac{\rho_0}{\zeta_0},
\end{align}
where Eqs.~\eqref{eq:rho0} and \eqref{eq:zeta0} were used in the last equality. In particular, the stationary toroidal spiral string with $n_1/n_2=\zeta_0/\rho_0=1$ is called the Hopf loop, which is discussed in Refs.~\cite{Igata:2009fd, Igata:2009dr}.

In addition to the Killing vector $\xi^\mu$, the vector
\begin{align}
\eta^\mu
=\partial_\sigma^\mu
=\alpha\partial^\mu_{\bar{\phi}}-\beta \partial^\mu_{\bar{\psi}},
\end{align}
which is tangent to a worldsheet of a toroidal spiral string, is a spacelike Killing vector. As a result, the worldsheet of the toroidal spiral string is flat because it is spanned by two commutable Killing vectors $\xi^\mu$ and $\eta^\mu$. Even if we consider a cohomogeneity-one string with a spacelike Killing vector, we can obtain a stationary toroidal string solution as is discussed in Refs.~\cite{Igata:2009fd, Igata:2009dr}.

\subsubsection{Planar Strings}
In this section, we discuss other special solutions with $L_1=L_2=0$. In this case, the potentials are in proportion to $\rho^2$ and $\zeta^2$ in Eqs.~\eqref{eq:rho_eq} and \eqref{eq:zeta_eq}, respectively. It is clear from Eqs.~\eqref{eq:phi_eq} and \eqref{eq:psi_eq} that we can choose solutions to be $\phi(\sigma)=\psi(\sigma)=0, \pi$ without loss of generality. In order to solve the residual equations, we introduce the rotating frame,
\begin{align}
(X, Y, Z, W)=(\rho \cos\phi, \rho \sin\phi, \zeta \cos\psi, \zeta \sin\psi).
\end{align}
Then Eqs.~\eqref{eq:rho_eq} and \eqref{eq:zeta_eq} are rewritten in the form
\begin{align}
&X'^2+\alpha^2 X^2=\frac12+K,\\
&Z'^2+\beta^2 Z^2=\frac12-K,
\end{align}
and $Y(\sigma)$ and $W(\sigma)$ vanish. Solutions for $X$ and $Z$ are given by
\begin{align}
X(\sigma)&=\frac{1}{\alpha}\sqrt{\frac12+K}\cos(\alpha \sigma),\\
Z(\sigma)&=\frac{1}{\beta}\sqrt{\frac12-K} \cos(\beta \sigma+\delta),
\end{align}
where $\delta$ is an integral constant.

\begin{figure}[htbp]
\bigskip
\begin{tabular}{cc}
\begin{minipage}{0.5\hsize}
\begin{center}
\includegraphics[width=6cm,clip]{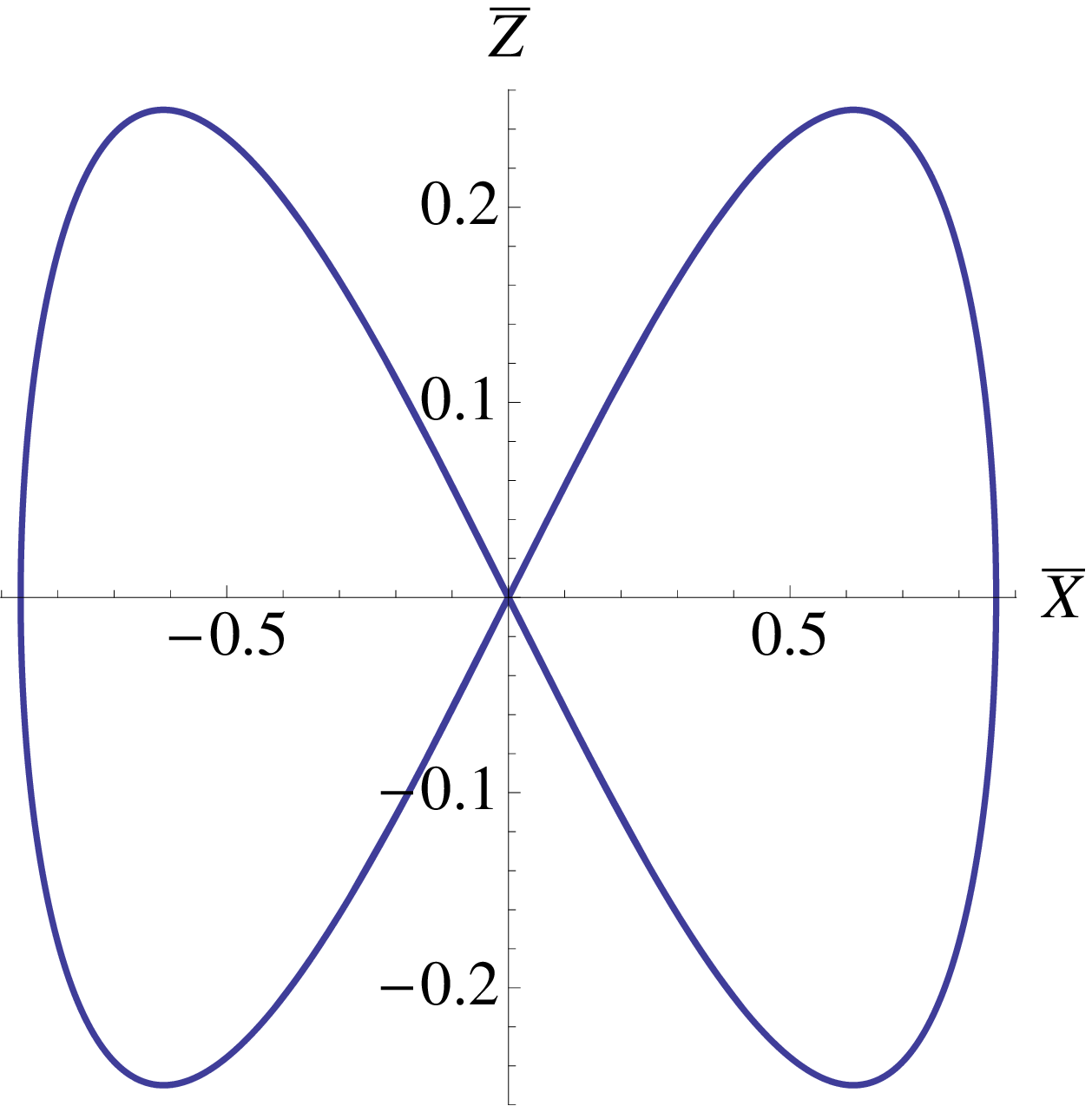}
\end{center}
\end{minipage}
\begin{minipage}{0.5\hsize}
\begin{center}
\includegraphics[width=6cm,clip]{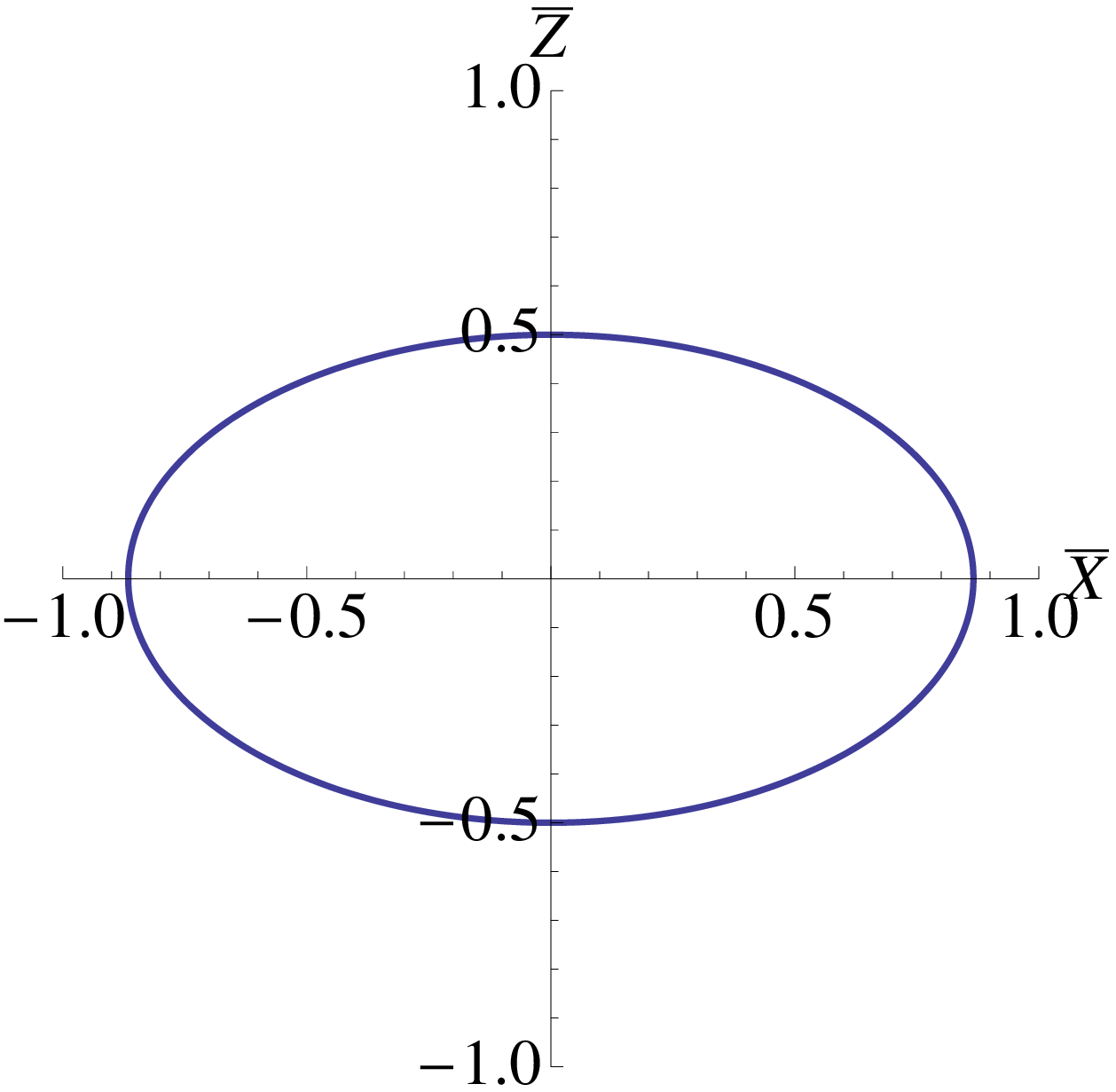}
\end{center}
\end{minipage}
\end{tabular}
\caption{The self intersecting planar string configuration with $\alpha=1$ and $\beta=2$ (left) and the planar loop with $\alpha=\beta=1$ (right) on the $\bar{X}$-$\bar{Z}$ plane at $\tau=0$ surface, where we have chosen $K=1/4$ and $\delta=\pi/2$. }
\label{fig:planar_string}
\bigskip
\end{figure}

In the Cartesian coordinates $\bar{T}$, $\bar{X}$, $\bar{Y}$, $\bar{Z}$, $\bar{W}$ in the five-dimensional flat spacetime, i.e., $ds^2=-d\bar{T}^2+d\bar{X}^2+d\bar{Y}^2+d\bar{Z}^2+d\bar{W}^2$, the solutions are represented as
\begin{align}
\bar{T}&=\tau,
\label{eq:static_gauge}\\
\bar{X}&=\frac{1}{\alpha}\sqrt{\frac12 +K} \cos(\alpha \tau)\cos(\alpha \sigma), \\
\bar{Y}&=\frac{1}{\alpha}\sqrt{\frac12 +K} \sin(\alpha \tau)\cos(\alpha \sigma), \\
\bar{Z}&=\frac{1}{\beta}\sqrt{\frac12 -K} \cos(\beta \tau)\cos(\beta \sigma+\delta), \\
\bar{W}&=\frac{1}{\beta}\sqrt{\frac12 -K} \sin(\beta \tau)\cos(\beta \sigma+\delta).
\label{eq:planar_sol}
\end{align}
At a moment $\bar{T}={\rm const.}$ we find $\bar{Y}/\bar{X}={\rm const.}$ and $\bar{W}/\bar{Z}={\rm const}$. Then the string described by \eqref{eq:static_gauge}--\eqref{eq:planar_sol} lies on the two-dimensional plane that is the intersection of two three-dimensional flat planes, $\bar{Y}/\bar{X}={\rm const.}$ and $\bar{W}/\bar{Z}={\rm const}$. Therefore, we call such a string as a planar string. The solutions show Lissajous figures on the two-dimensional plane, as illustrated in Fig.~\ref{fig:planar_string}. While the solutions have self intersecting points in the case $\alpha\neq\beta$, the string configuration in the special case $\alpha=\beta$ is described by an oval on the two-dimensional plane, as illustrated in Fig.~\ref{fig:planar_string}. We call the string the planar loop.
The parameters $K$, $\delta$ are a measure of ellipticity of the planar loop and inclination of the major axis. Because we can adjust the inclination angle to zero by coordinate rotation, we can fix $\delta=\pi/2$ without loss of generality, and then the planar loop solution satisfies
\begin{align}
\frac{\bar{X}^2+\bar{Y}^2}{1/2+K}+\frac{\bar{Z}^2+\bar{W}^2}{1/2-K}=\alpha^{-2}.
\label{eq:ellipsoid}
\end{align}
That is, the planar loop lies on the ellipsoid~\eqref{eq:ellipsoid}. In particular, in the case $K=0$ the planar loop lies on the three-dimensional round sphere and we represent the solution in new coordinates
\begin{align}
\bar{X}_\pm&=\frac{\bar{X}\pm \bar{W}}{\sqrt{2}}=\frac{1}{2\alpha}\cos \alpha(\sigma\pm\tau), \\
\bar{Z}_\pm&=\frac{-\bar{Z}\pm \bar{Y}}{\sqrt{2}}=\pm\frac{1}{2\alpha}\sin \alpha(\sigma\pm\tau),
\end{align}
where the spacetime metric takes the form $ds^2=-dT^2+dX_+^2+dZ_+^2+dX_-^2+dZ_-^2$. This solution reproduces the Hopf loop solution discussed at the end of the previous section.

\subsection{General Cases}
Now, we discuss generic configuration of stationary rotating closed strings. The general solutions to Eqs.~\eqref{eq:rho_eq}--\eqref{eq:psi_eq} are written as
\begin{align}
\rho^2(\sigma)&=\frac{\rho^2_+-\rho^2_-}{2}\cos\left(2\alpha \sigma\right)+\frac{\rho^2_++\rho^2_-}{2},
\label{eq:rho}\\
\zeta^2(\sigma)&=\frac{\zeta^2_+-\zeta^2_-}{2}\cos\left[2(\beta \sigma + \delta)\right]+\frac{\zeta^2_++\zeta^2_-}{2},
\label{eq:zeta}\\
\phi(\sigma)&=\mathrm{sgn}(l_1)\left(\arctan \left(\frac{\rho_-}{\rho_+}\tan(\alpha\sigma)\right)+\pi\left\lfloor\frac{\alpha\sigma+\pi/2}{\pi}\right\rfloor \right)-\alpha C\sigma,
\label{eq:phi}\\
\psi(\sigma)&=\mathrm{sgn}(l_2)\left(\arctan \left(\frac{\zeta_-}{\zeta_+}\tan(\beta\sigma+\delta)\right)+\pi\left\lfloor \frac{\beta \sigma+\delta+\pi/2}{\pi}\right\rfloor\right)-\beta C\sigma,
\label{eq:psi}
\end{align}
where $\delta$ is an integral constant, $\rho_\pm, \zeta_\pm$ are given by Eqs.~\eqref{eq:rho_pm} and \eqref{eq:zeta_pm}, and $\lfloor x \rfloor$ and $\mathrm{sgn}(x)$ denote the floor function and the sign function of $x$, respectively.

By periodicity of $\rho$ and $\zeta$ in Eqs.~\eqref{eq:rho} and \eqref{eq:zeta}, the solutions satisfy
\begin{align}
&\rho(\sigma+\sigma_{{\rm p}})=\rho(\sigma),\\
&\zeta(\sigma+\bar{\sigma}_{{\rm p}}) = \zeta(\sigma),\\
&\phi(\sigma+k\sigma_{{\rm p}})=\phi(\sigma)+\pi k\left(\mathrm{sgn}(l_1)-C\right),\\
&\psi(\sigma+\bar k \bar{\sigma}_{{\rm p}})=\psi(\sigma)+\pi\bar k\left(\mathrm{sgn}(l_2)-C\right),
\end{align}
where $\sigma_{{\rm p}}=\pi/\alpha, \bar{\sigma}_{{\rm p}}=\pi/\beta$, and $k, \bar k$ are natural numbers.

In order that projections of orbits on the $\rho$-$\phi$ plane and the $\zeta$-$\psi$ plane are closed, the constant $C$ must satisfy
\begin{align}
&\pi k\left(\mathrm{sgn}(l_1)-C\right)=2\pi m,
\label{eq:XYclosed}\\
&\pi \bar k\left(\mathrm{sgn}(l_2)-C\right)=2\pi\bar m,
\label{eq:ZWclosed}
\end{align}
where $m$ and $\bar m$ are integers.  We can assume $(k, |m|)$ and $(\bar k, |\bar m|)$ are pairs of relatively primes.

From Eqs.~\eqref{eq:C_range}, \eqref{eq:XYclosed}, and \eqref{eq:ZWclosed}, we find that $C$ is a rational number in the range $-1<C<1$. Since we can assume $l_1>0$ without loss of generality, we consider the following three cases: $l_2>0, l_2=0$, and $l_2<0$, separately.

In the case $l_1>0$ and $l_2>0$, we see $k=\bar k$, $m=\bar m$, and $C$ is in the range $0<C<1$, and then we have
\begin{align}
0<m=\bar m<\frac{k}{2}=\frac{\bar k}{2}.
\end{align}
In the case $l_1>0$ and $l_2<0$, we see $k=\bar k$, $m-k=\bar m$, and $C$ can be in the range $-1<C<1$, and then we have
\begin{align}
0<m<k, \quad \mbox{and}\quad -\bar k<\bar m<0.
\end{align}
Examples of possible combinations of $(k,m;\bar k, \bar m)$ are shown in Tables \ref{table:1} and \ref{table:2}. As illustrated in Figs.~\ref{fig:general_loop+} for $l_2>0$ case, and in Figs.~\ref{fig:general_loop-} for $l_2<0$ case, the projected closed orbits on the $\rho$-$\phi$ plane are rounded polygons or rounded star polygons, where each segment is a curve starting from $\rho_+$ through $\rho_-$ and ending at $\rho_+$. The rounded (star) polygon with $(k, m)$ has $k$ segments, and the closed curve wraps the center of two-dimensional plane $m$ times. The projected closed orbits on the $\zeta$-$\psi$ plane with $(\bar k, \bar m)$ have the same properties.

The limits $l_2\to 0+$ in the positive $l_2$ case and $l_2\to 0-$ in the negative $l_2$ case give the same solutions for $l_2=0$. Projected closed orbits are shown in Fig.~\ref{fig:general_loop_zerol}, where the curves on the $\zeta$-$\psi$ plane pass through the central point $\zeta=0$.

In order that the string is closed, we should additionally require the condition
\begin{align}
n k \sigma_{\rm p}=\bar n \bar k \bar{\sigma}_{\rm p},
\label{eq:loop_condition}
\end{align}
where $n$ and $\bar n$ are relatively prime. On a closed curve of the string, the pattern of rounded (star) polygon with $(k, m)$ appears $n$ times while the rounded (star) polygon with $(\bar k, \bar m)$ appears $\bar n$ times. Since $k=\bar k$ we find from Eq.~\eqref{eq:loop_condition} that
\begin{equation}
\frac{\beta}{\alpha}=\frac{\bar n}{n},
\label{eq:winding_ratio}
\end{equation}
then the ratio $\beta/\alpha$ is a rational number.

\begin{table}[t]
\bigskip
\begin{tabular}{lllll}
\hline\hline
&\multicolumn{2}{l}{($k,m; \bar k,\bar m$)}&&\\
\hline
&$(3,1; 3,1)$&&&\\
&$(4,1; 4,1)$&&&\\
&$(5,1; 5,1)$,&$(5,2; 5,2)$&&\\
&$(6,1; 6,1)$&&&\\
&$(7,1; 7,1)$,&$(7,2; 7,2)$,&$(7,3; 7,3)$&\\
&$(8,1; 8,1)$,&$(8,3; 8,3)$&&\\
&$\ \cdots$& &&\\
\hline\hline
\end{tabular}
\caption{Possible combinations of $(k, m; \bar k, \bar m)$ for $k\leq 8$ in the case $l_1>0$ and $l_2>0$. The value of $C$ is given by $C=1-2m/k$. }
\label{table:1}
\end{table}
\bigskip
{~}
\begin{table}[t]
\begin{tabular}{lllllll}
\hline\hline
&($k,m; \bar k, \bar m$)&&\\
\hline
&$(2,1; 2,-1)$&&&\\
&$(3,1; 3,-2)$&&&\\
&$(4,1; 4,-3)$&&&\\
&$(5,1; 5,-4)$,&$(5,2; 5,-3)$,&$(5,3; 5,-2)$,&$(5,4; 5,-1)$&\\
&$(6,1; 6,-5)$,&$(6,5; 6,-1)$\\
&$(7,1; 7,-6)$,&$(7,2; 7, -5)$,&$(7,3; 7, -4)$,&$(7,4; 7,-3)$,&$(7,5; 7, -2)$,&$(7,6; 7, -1)$\\
&$(8,1; 8,-7)$,&$(8,3; 8, -5)$,&$(8,5; 8, -3)$,&$(8,7; 8, -1)$\\
&$\ \cdots$& &&\\
\hline\hline
\end{tabular}
\caption{Possible combinations of $(k, m; \bar k, \bar m)$ for $k\leq 8$ in the case $l_1>0$ and $l_2<0$. The value of $C$ is given by $C=1-2m/k=-1-2\bar m/\bar k$. }
\label{table:2}
\medskip
\end{table}

\newpage
\def\figsize{4.20cm}

\begin{figure}[!h]
\bigskip
\bigskip
\begin{tabular}{ccc}
\begin{minipage}[]{0.33\hsize}
\begin{tabular}{c}
\begin{minipage}[]{1\hsize}
\begin{center}
\includegraphics[width=\figsize,clip]{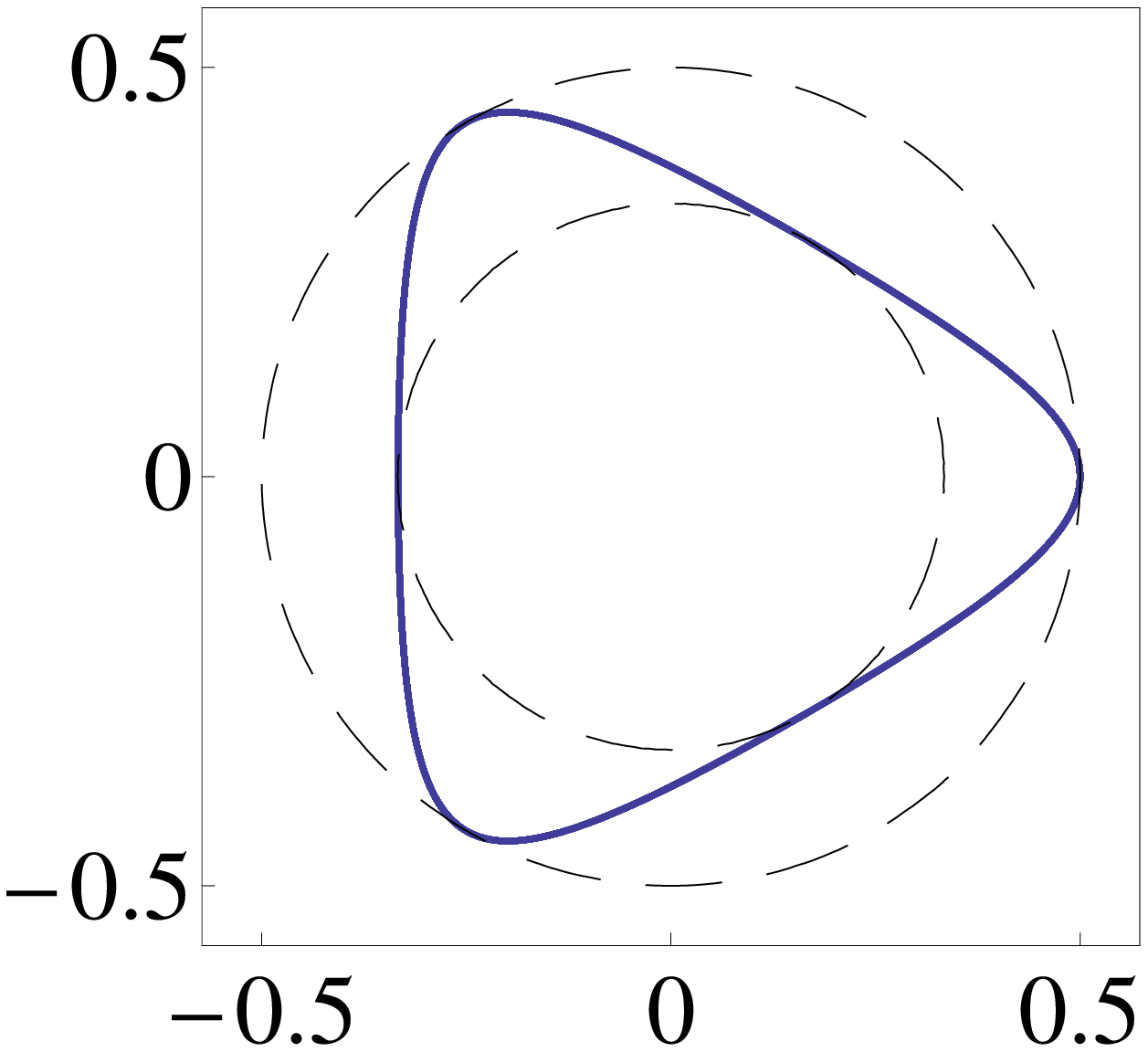}
\end{center}
\end{minipage}
\\
\begin{minipage}[]{1\hsize}
\begin{center}
\includegraphics[width=\figsize,clip]{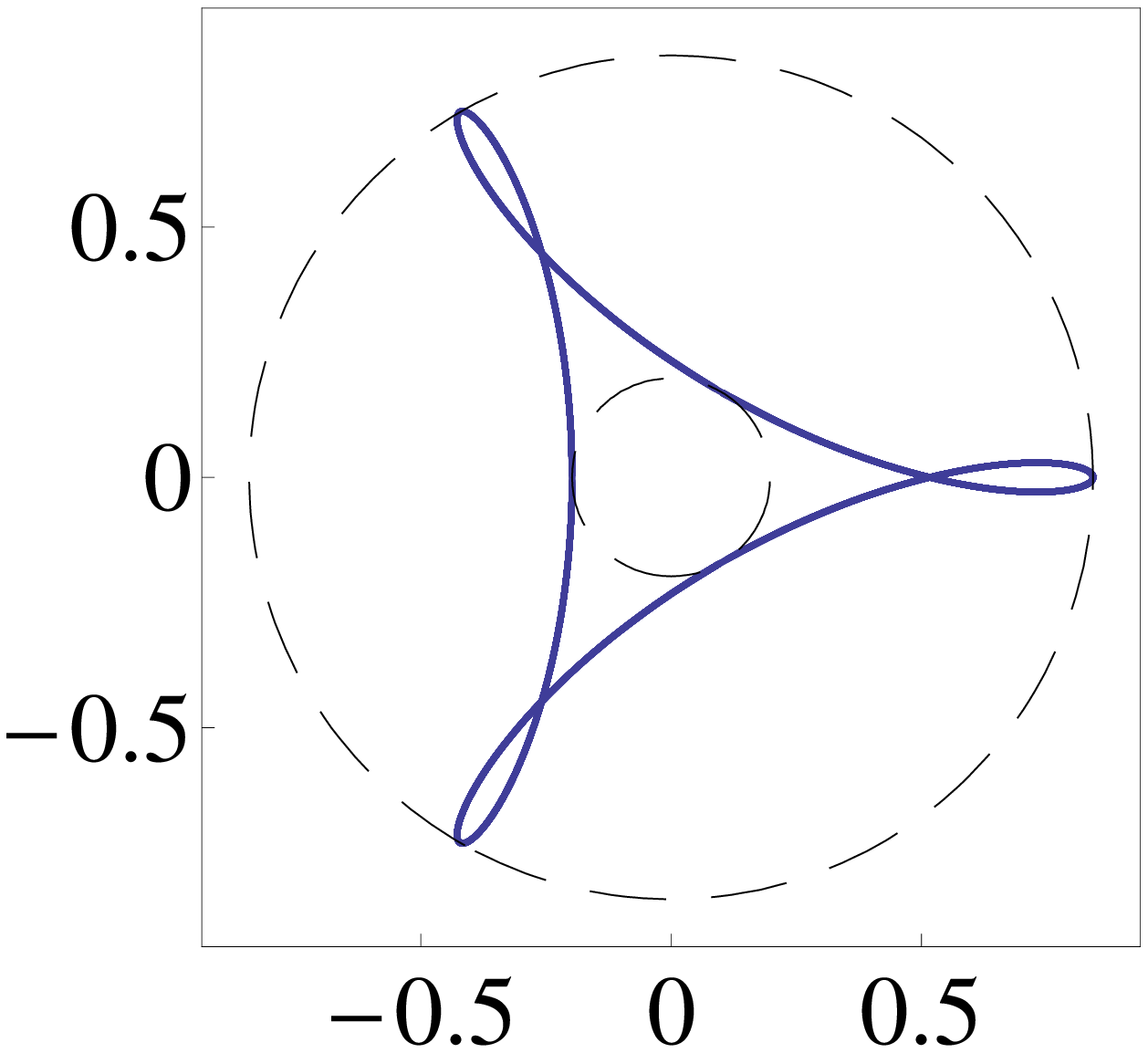}
\end{center}
\end{minipage}
\end{tabular}
\begin{flushleft}
\qquad \quad (a) $l_1=1/6,~l_2=1/6$, \\
\qquad \qquad$(k,m; \bar k, \bar m)=(3,1; 3,1)$
\end{flushleft}
\end{minipage}
\begin{minipage}[]{0.33\hsize}
\begin{tabular}{c}
\begin{minipage}[]{1\hsize}
\begin{center}
\includegraphics[width=\figsize,clip]{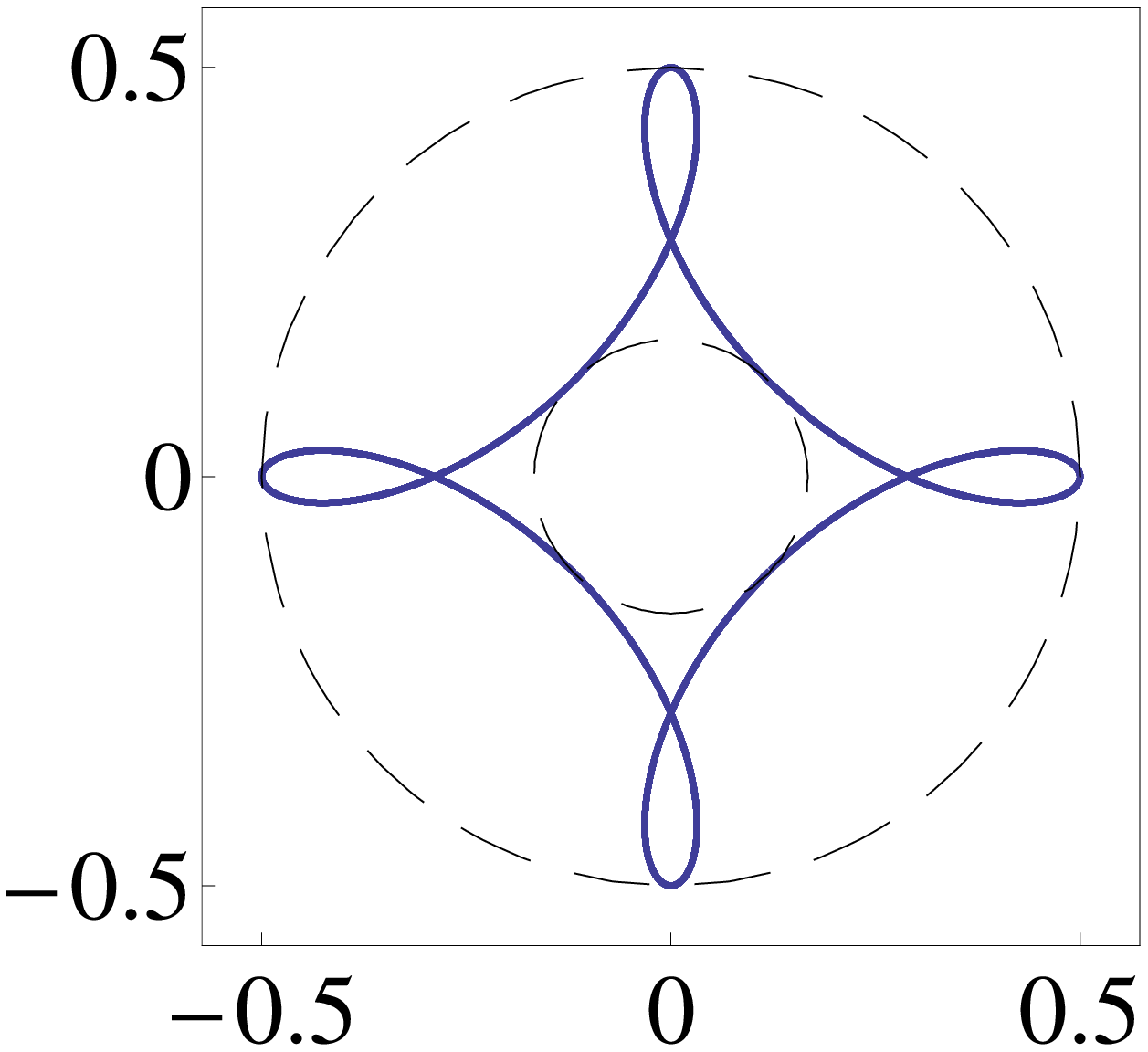}
\end{center}
\end{minipage}
\\
\begin{minipage}[]{1\hsize}
\begin{center}
\includegraphics[width=\figsize,clip]{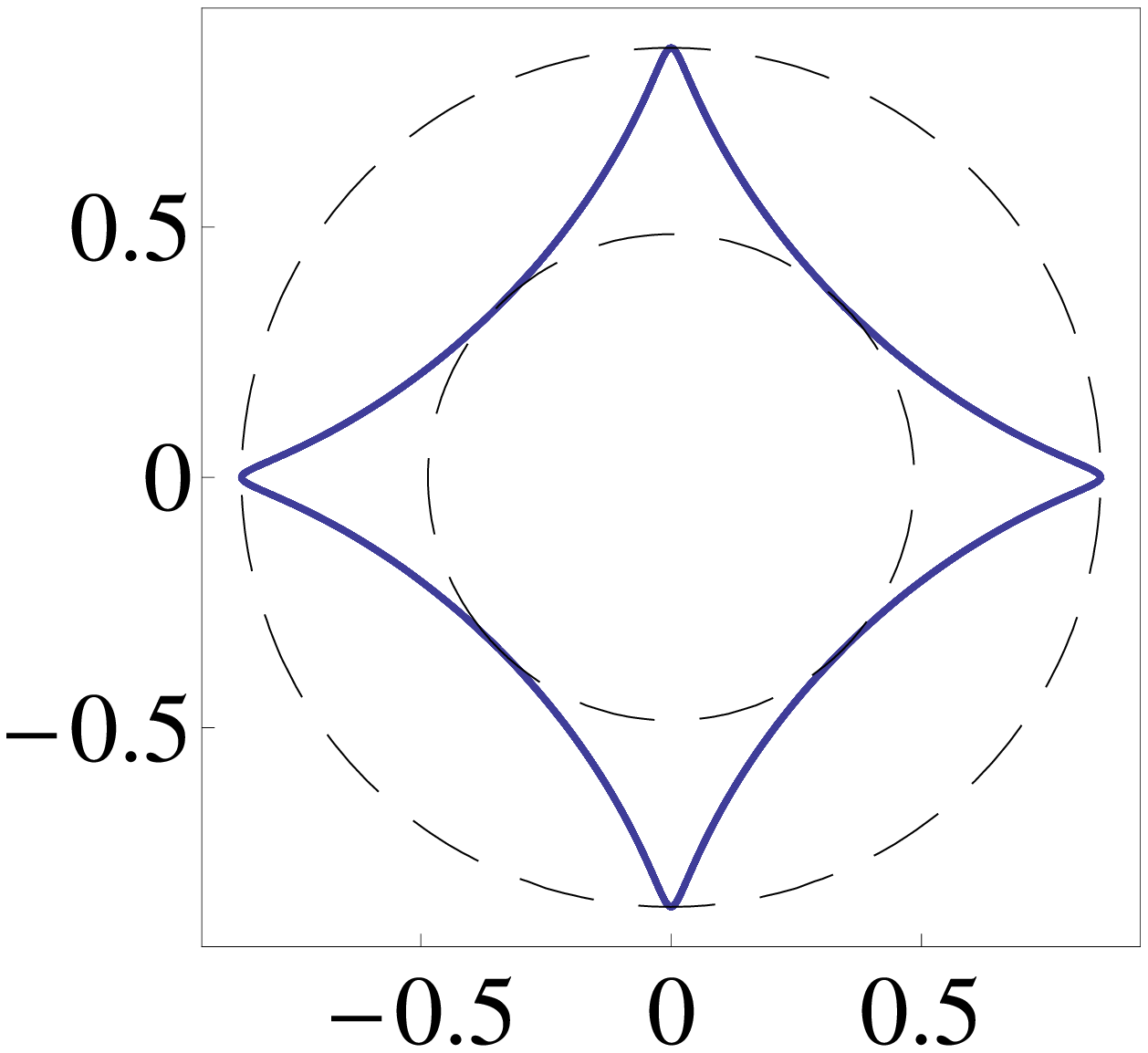}
\end{center}
\end{minipage}
\end{tabular}
\begin{flushleft}
\qquad \quad (b) $l_1=1/12,~l_2=5/12$, \\
\qquad \qquad$(k,m; \bar k, \bar m)=(4,1; 4,1)$
\end{flushleft}
\end{minipage}
\begin{minipage}[]{0.33\hsize}
\begin{tabular}{c}
\begin{minipage}[]{1\hsize}
\begin{center}
\includegraphics[width=\figsize,clip]{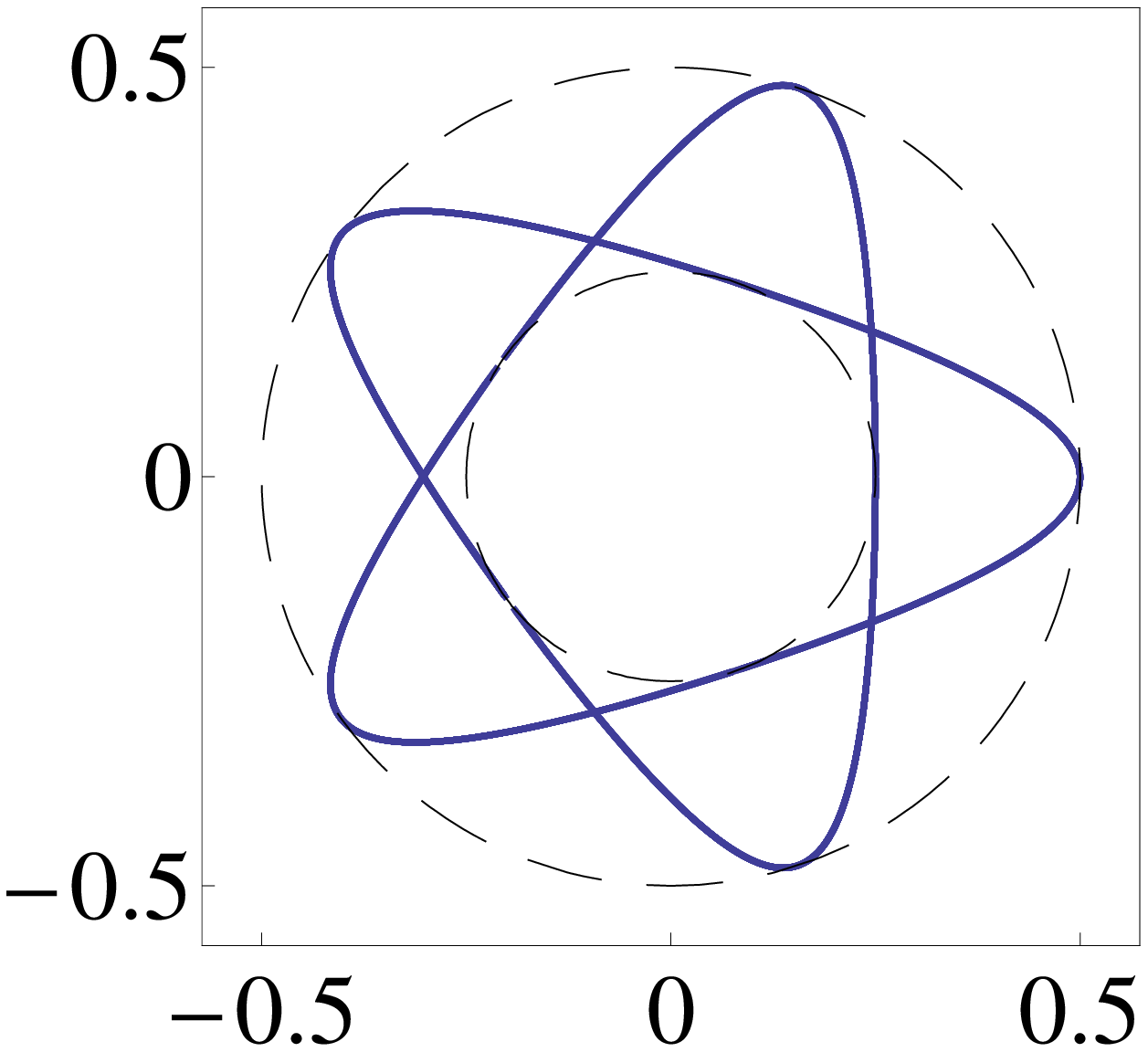}
\end{center}
\end{minipage}
\\
\begin{minipage}[]{1\hsize}
\begin{center}
\includegraphics[width=\figsize,clip]{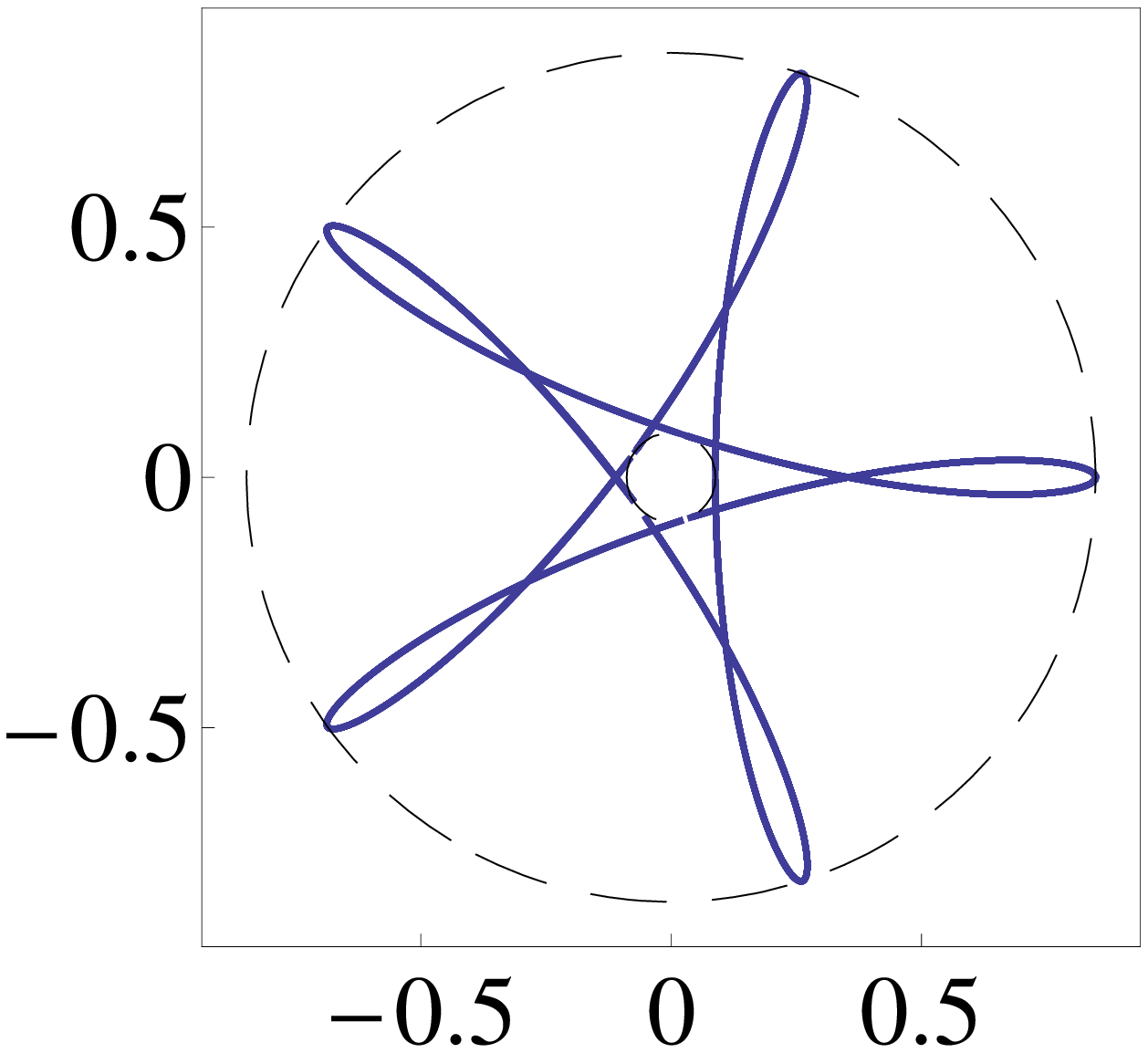}
\end{center}
\end{minipage}
\end{tabular}
\begin{flushleft}
\qquad \quad (c) $l_1=1/8,~l_2=3/40$, \\
\qquad \qquad $(k,m; \bar k,\bar m)=(5,2; 5,2)$
\end{flushleft}
\end{minipage}
\end{tabular}
\\
\bigskip
\bigskip
\begin{tabular}{ccc}
\begin{minipage}[]{0.33\hsize}
\begin{tabular}{c}
\begin{minipage}[]{1\hsize}
\begin{center}
\includegraphics[width=\figsize,clip]{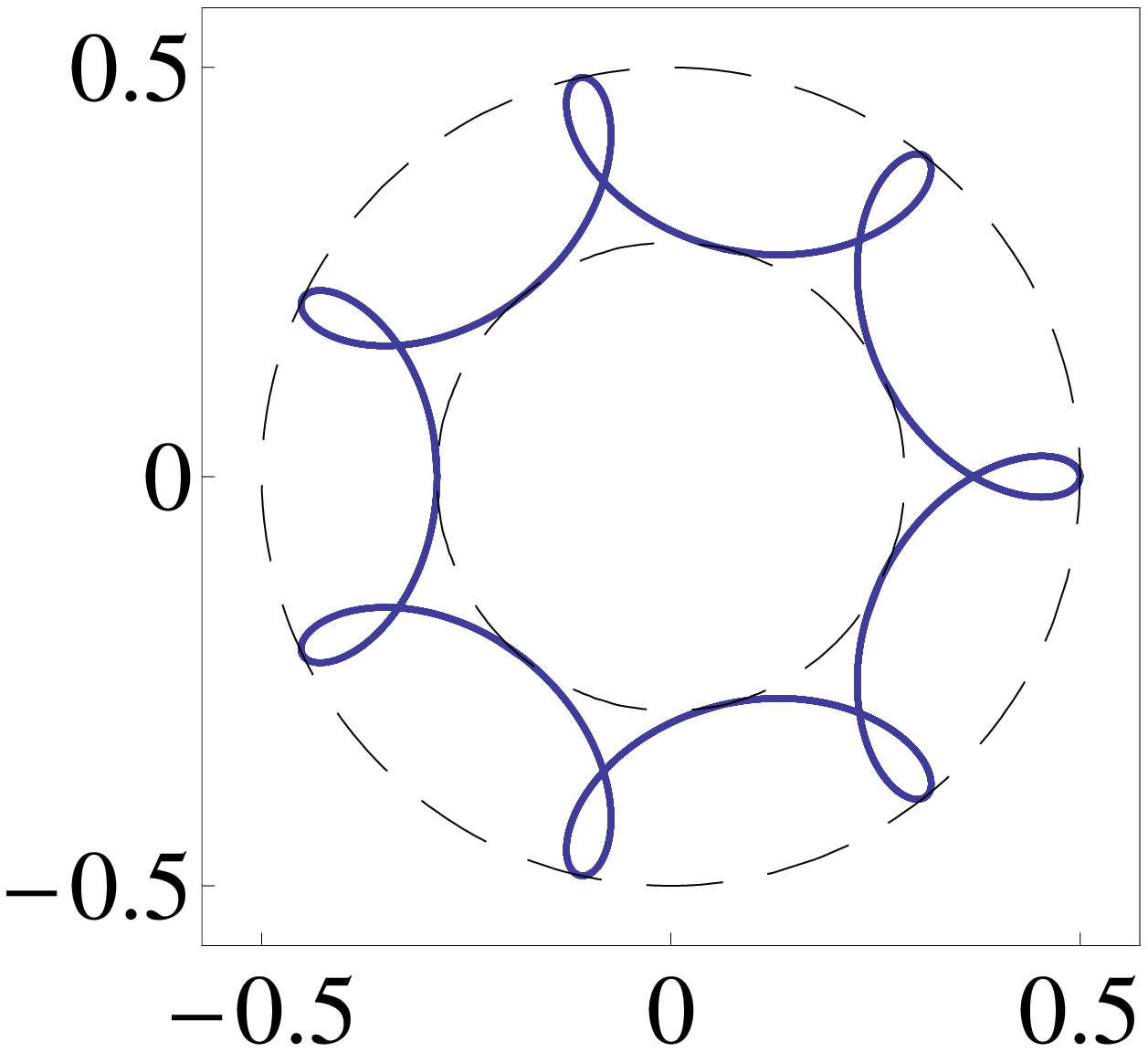}
\end{center}
\end{minipage}
\\
\begin{minipage}[]{1\hsize}
\begin{center}
\includegraphics[width=\figsize,clip]{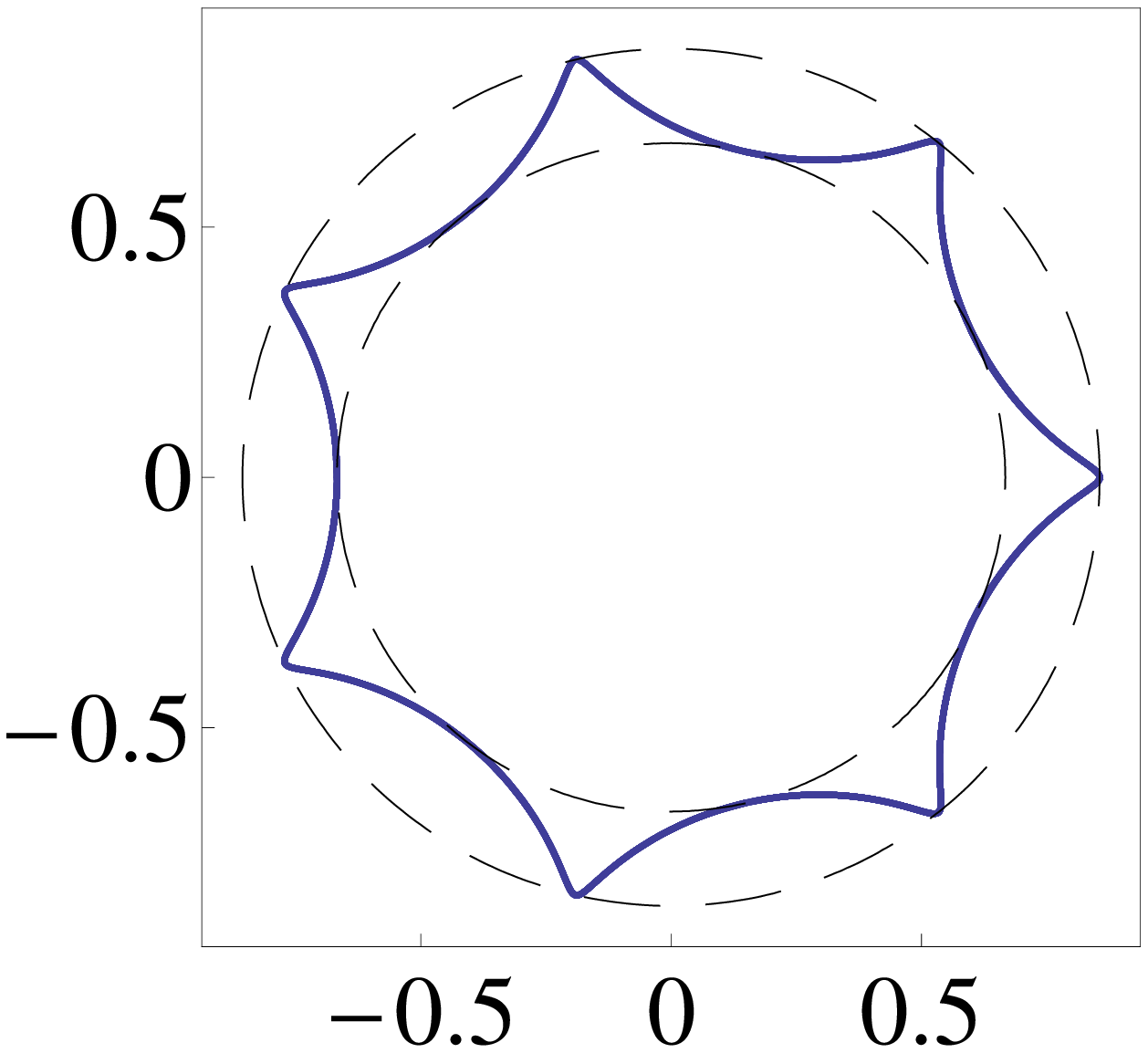}
\end{center}
\end{minipage}
\end{tabular}
\begin{flushleft}
\qquad \quad (d) $l_1=1/7,~l_2=4/7$, \\
\qquad \qquad $(k,m; \bar k,\bar m)=(7,1; 7,1)$
\end{flushleft}
\end{minipage}
\begin{minipage}[]{0.33\hsize}
\begin{tabular}{c}
\begin{minipage}[]{1\hsize}
\begin{center}
\includegraphics[width=\figsize,clip]{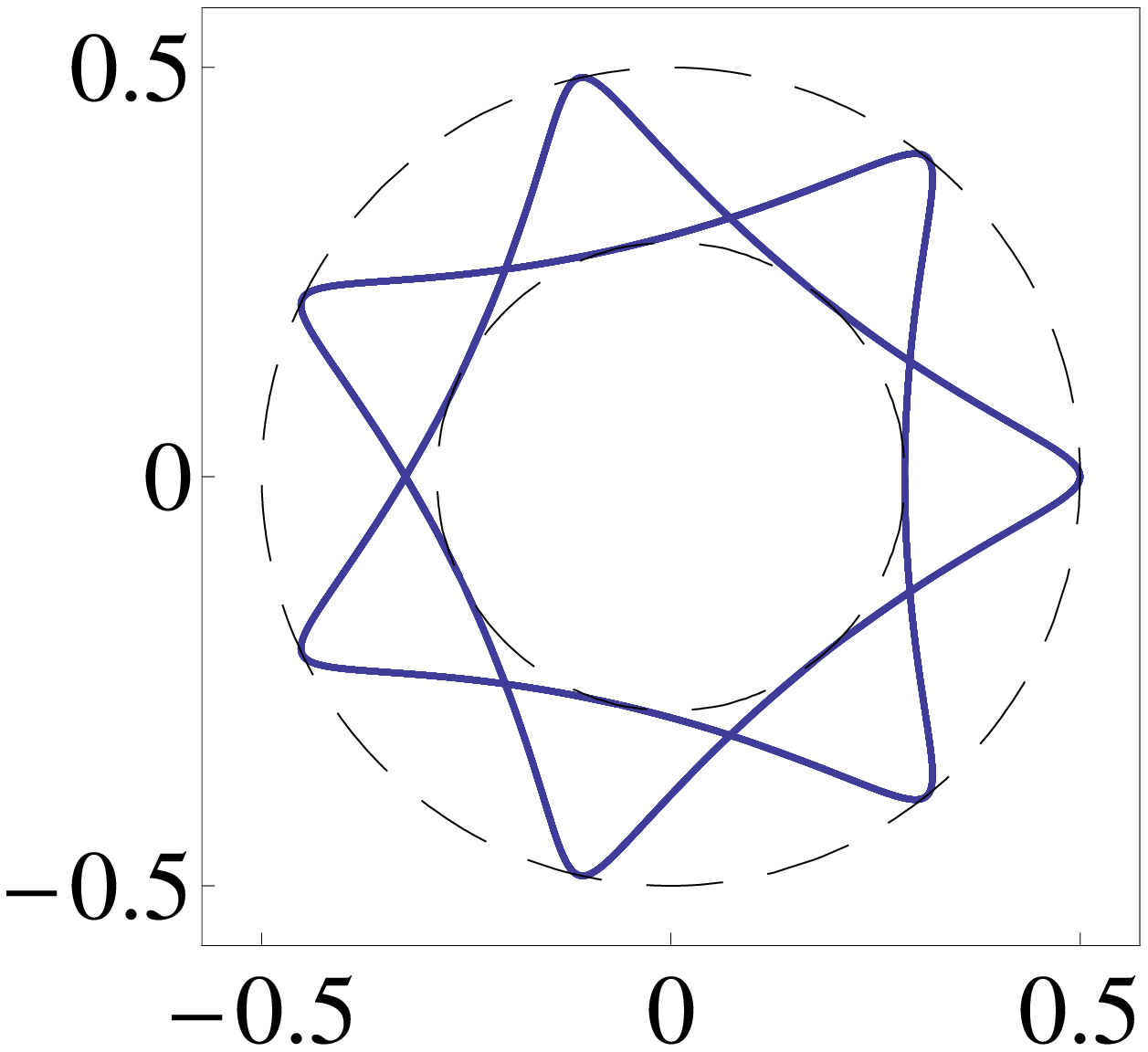}
\end{center}
\end{minipage}
\\
\begin{minipage}[]{1\hsize}
\begin{center}
\includegraphics[width=\figsize,clip]{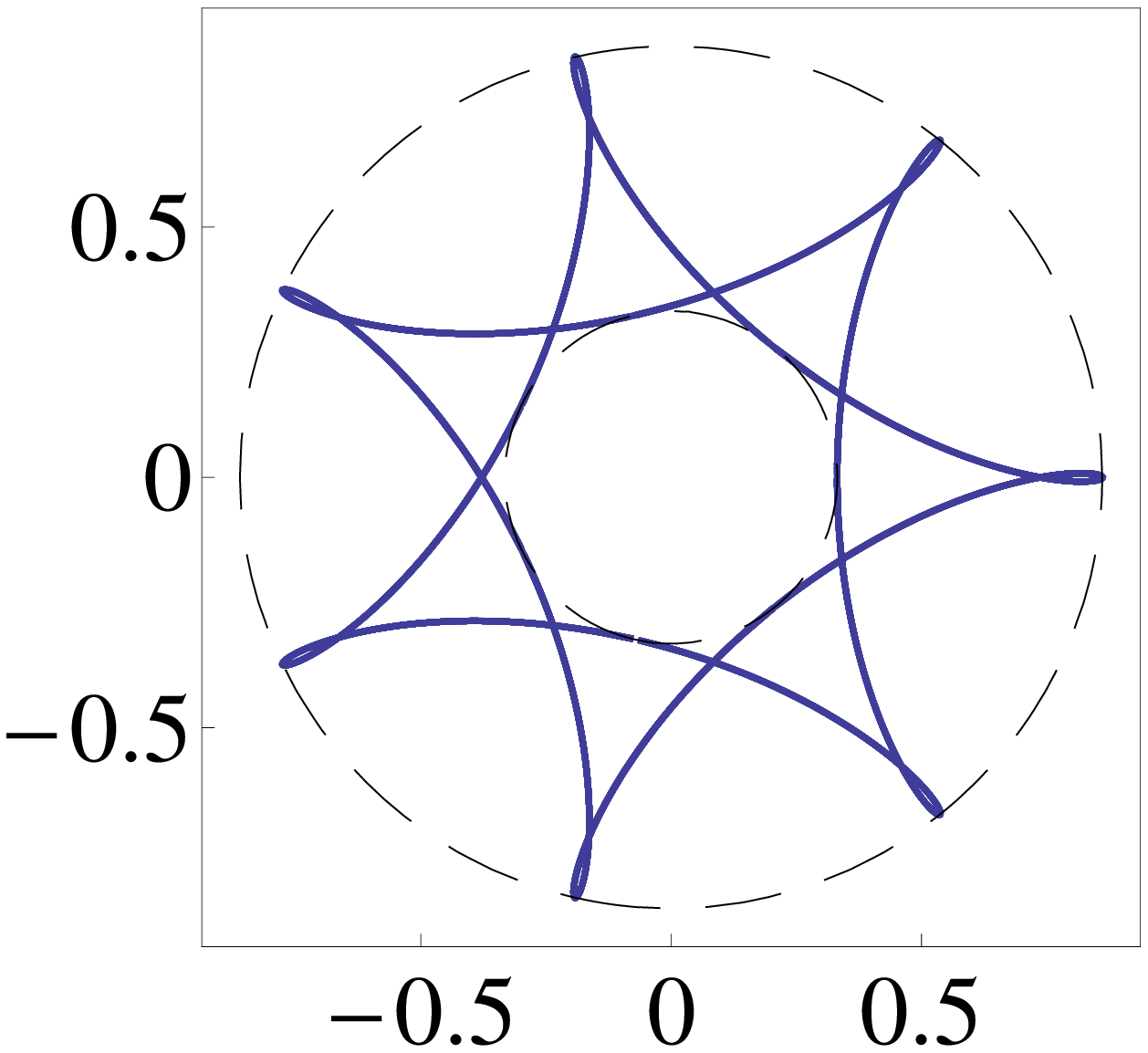}
\end{center}
\end{minipage}
\end{tabular}
\begin{flushleft}
\qquad \quad (e) $l_1=1/7,~l_2=2/7$, \\
\qquad \qquad $(k,m; \bar k,\bar m)=(7,2; 7,2)$
\end{flushleft}
\end{minipage}
\begin{minipage}[]{0.33\hsize}
\begin{tabular}{c}
\begin{minipage}[]{1\hsize}
\begin{center}
\includegraphics[width=\figsize,clip]{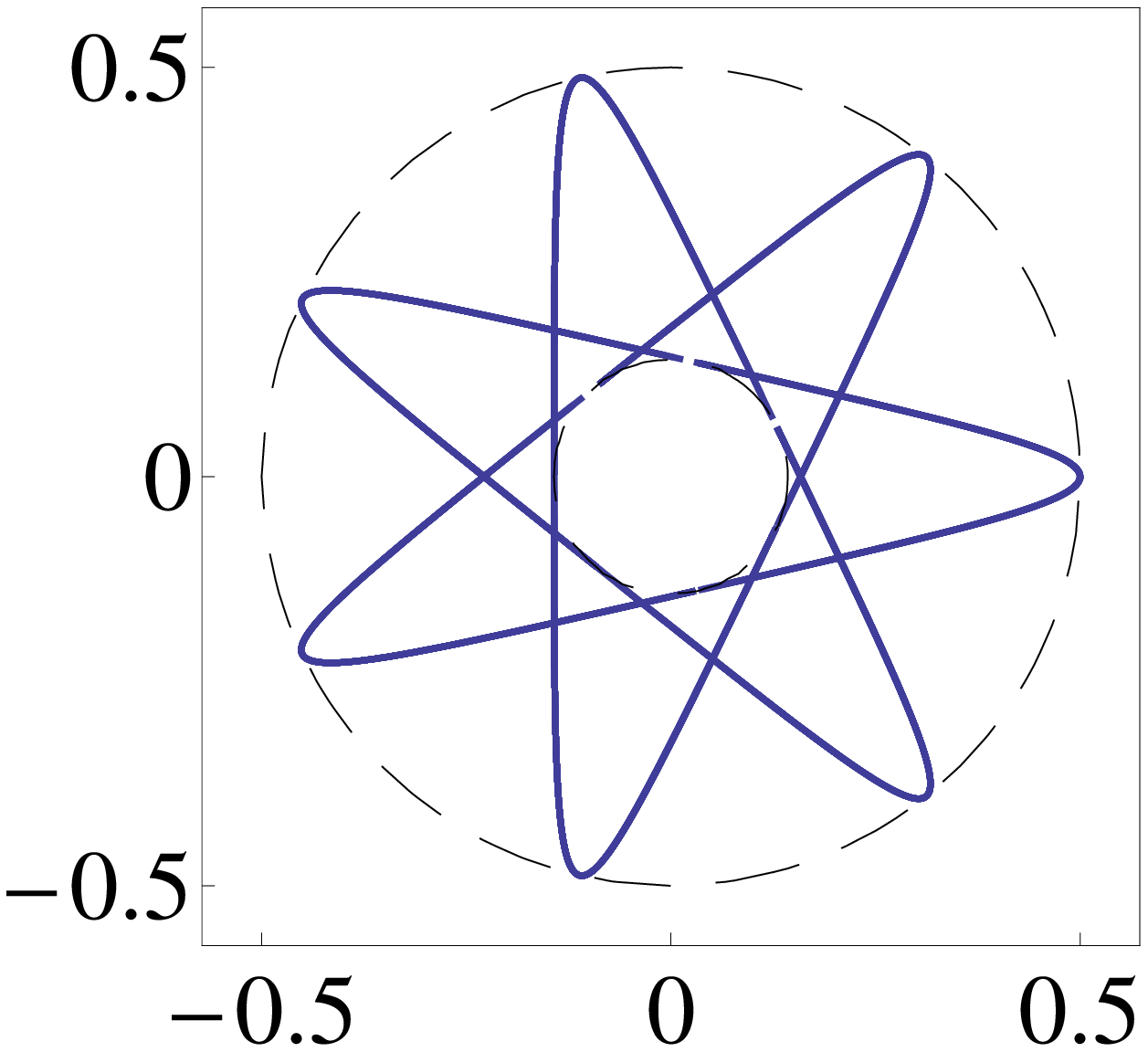}
\end{center}
\end{minipage}
\\
\begin{minipage}[]{1\hsize}
\begin{center}
\includegraphics[width=\figsize,clip]{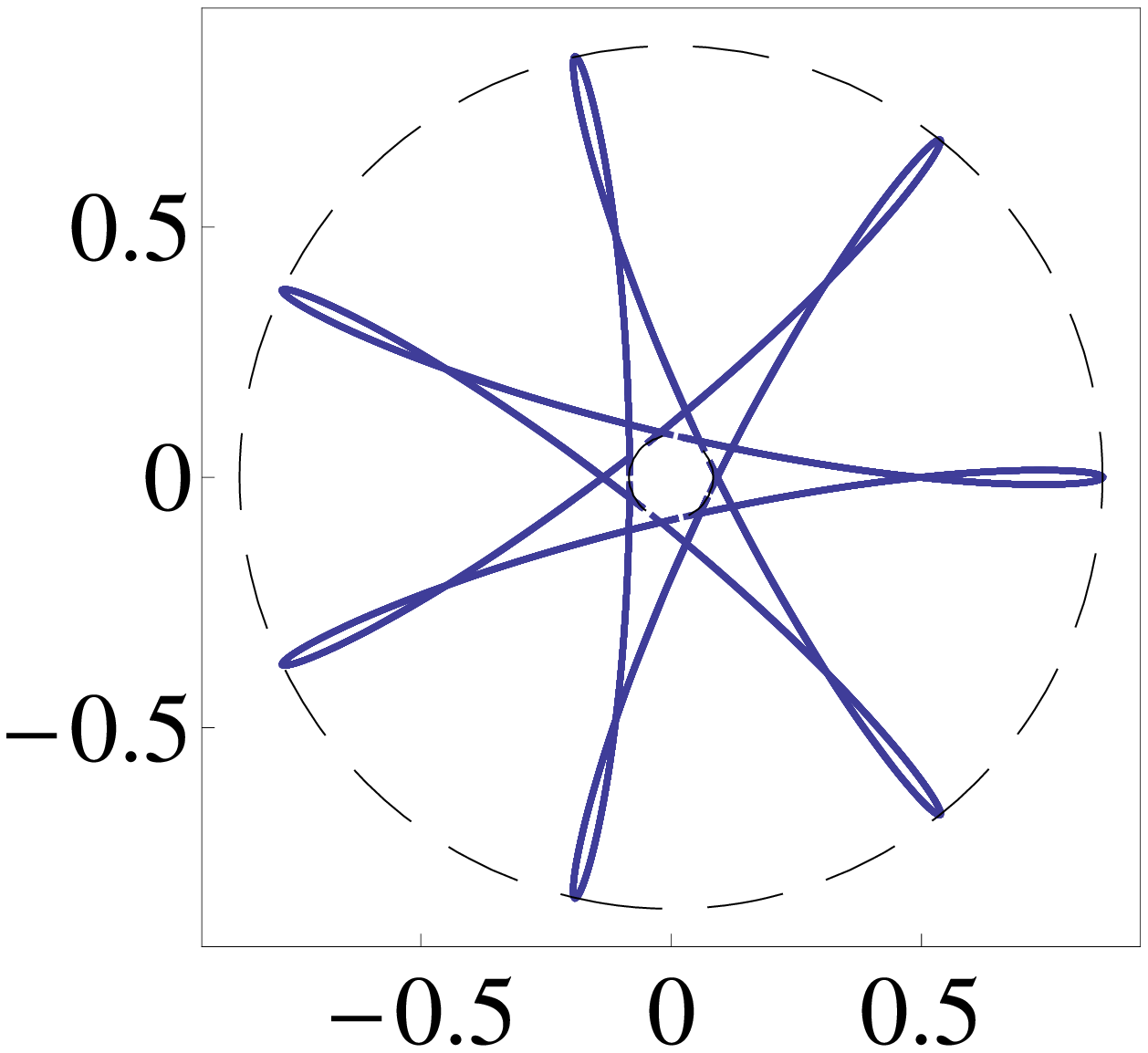}
\end{center}
\end{minipage}
\end{tabular}
\begin{flushleft}
\qquad \quad (f) $l_1=1/14,~l_2=1/14$, \\
\qquad \qquad $(k,m; \bar k,\bar m)=(7,3; 7,3)$
\end{flushleft}
\end{minipage}
\end{tabular}
\smallskip
\caption{
The projected closed orbits (a)--(f) for various parameters $l_1>0,~l_2>0,~(k,m; \bar k, \bar m)$ on $\rho$-$\phi$ plane (upper panels) and $\zeta$-$\psi$ plane (lower panels). The radius $\rho_+$ is normalized to $\rho_+=1/2$, and we choose $\alpha=\beta=1$. Dashed circles show $\rho_+,\rho_-$ in the upper panels, and $\zeta_+, \zeta_-$ in the lower panels. }
\label{fig:general_loop+}
\end{figure}

\newpage
\begin{figure}[!h]
\bigskip
\bigskip
\begin{tabular}{ccc}
\begin{minipage}[]{0.33\hsize}
\begin{tabular}{c}
\begin{minipage}[]{1\hsize}
\begin{center}
\includegraphics[width=\figsize,clip]{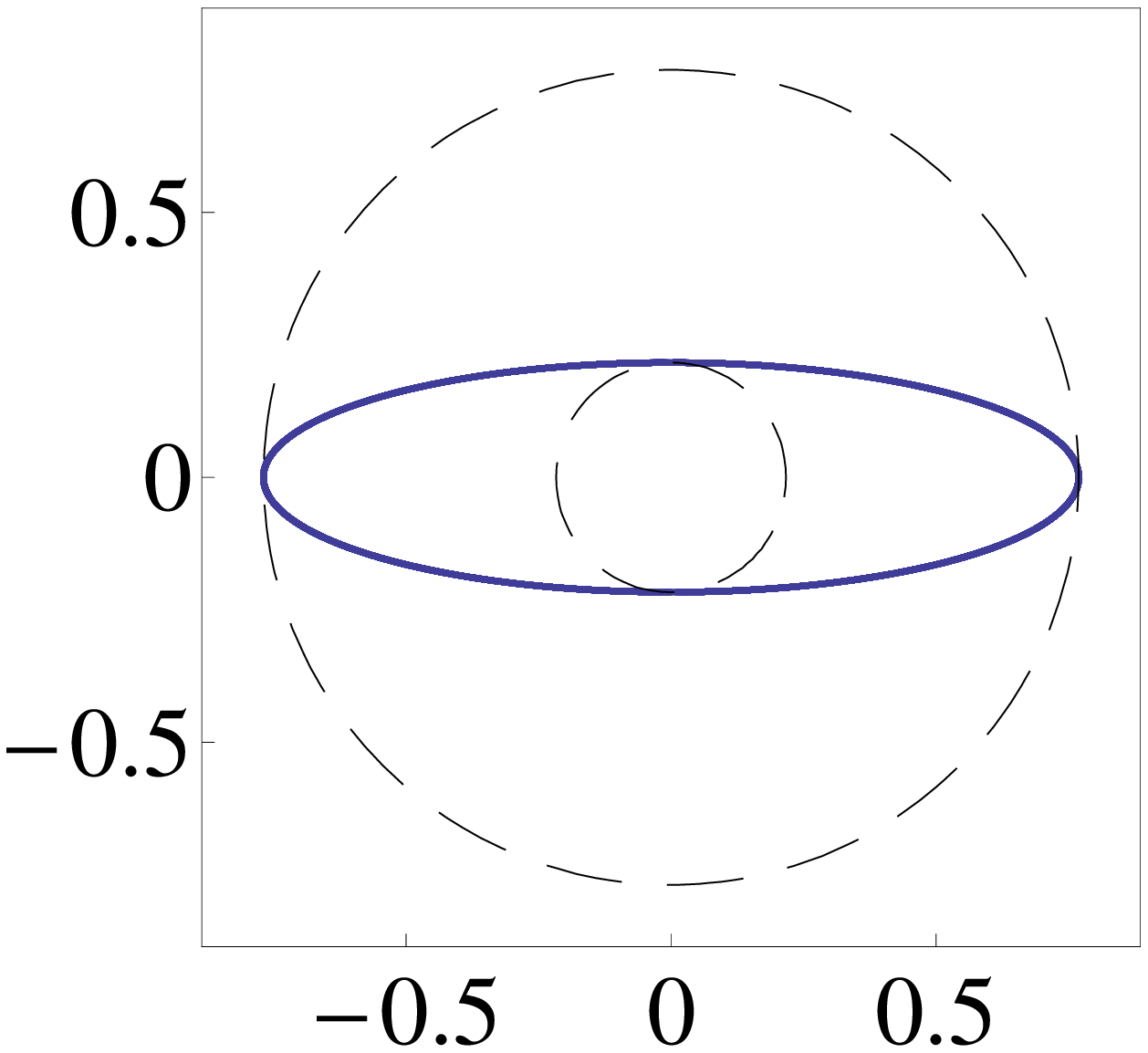}
\end{center}
\end{minipage}
\\
\begin{minipage}[]{1\hsize}
\begin{center}
\includegraphics[width=\figsize,clip]{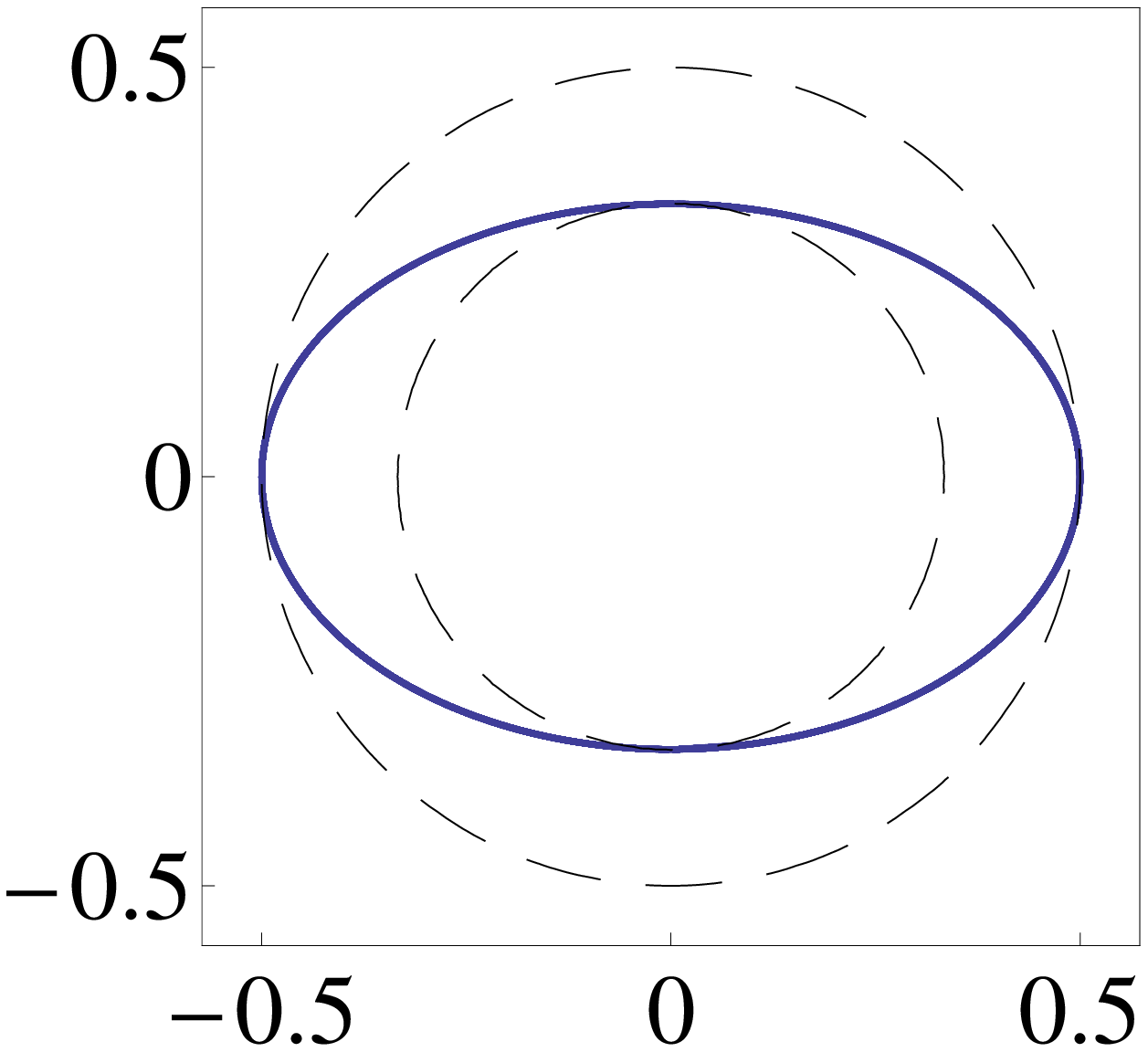}
\end{center}
\end{minipage}
\end{tabular}
\begin{flushleft}
\qquad \quad (a) $l_1=1/6,~l_2=-1/6$, \\
\qquad \qquad $(k,m; \bar k,\bar m)=(2,1; 2,-1)$
\end{flushleft}
\end{minipage}
\begin{minipage}[]{0.33\hsize}
\begin{tabular}{c}
\begin{minipage}[]{1\hsize}
\begin{center}
\includegraphics[width=\figsize,clip]{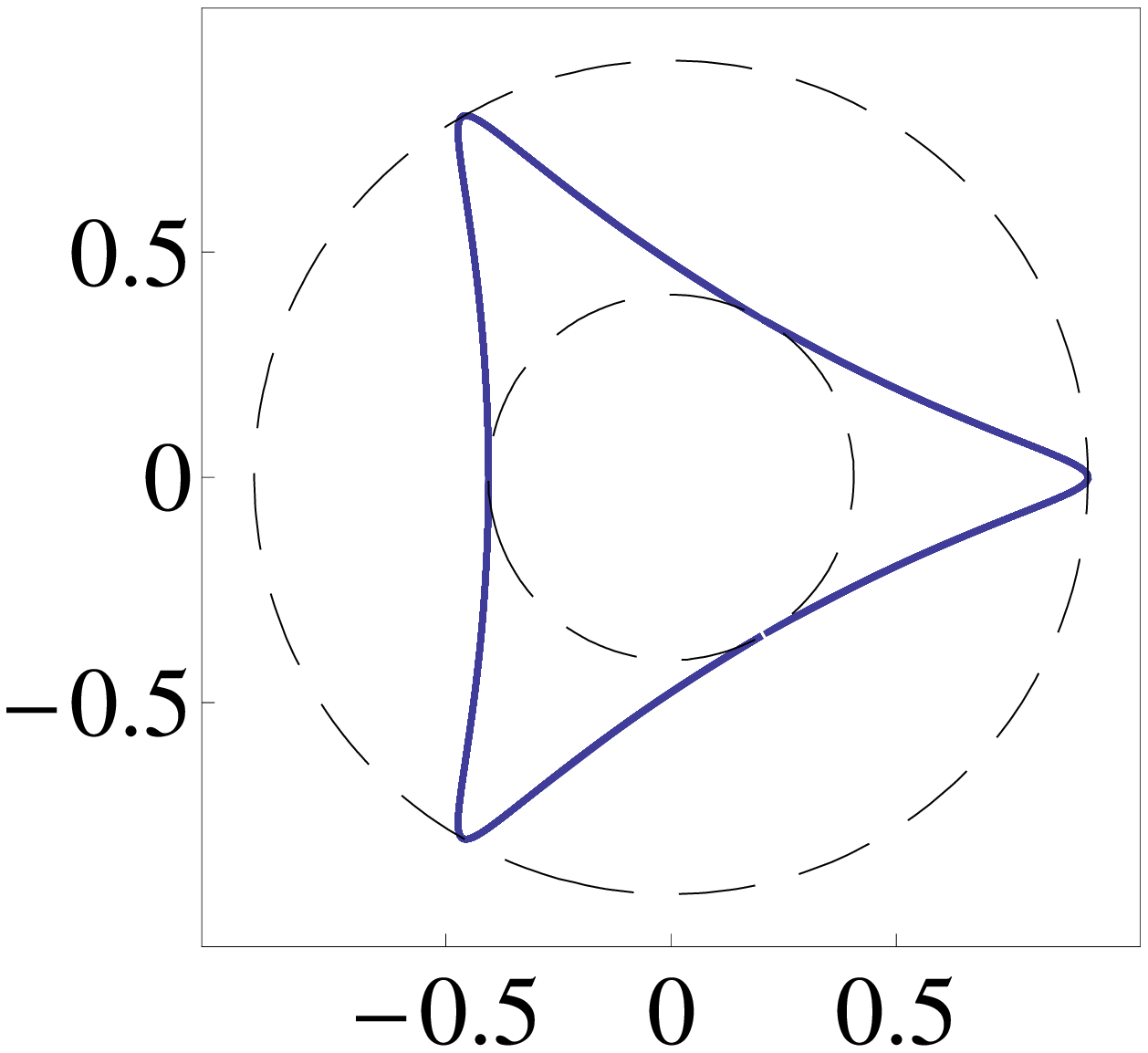}
\end{center}
\end{minipage}
\\
\begin{minipage}[]{1\hsize}
\begin{center}
\includegraphics[width=\figsize,clip]{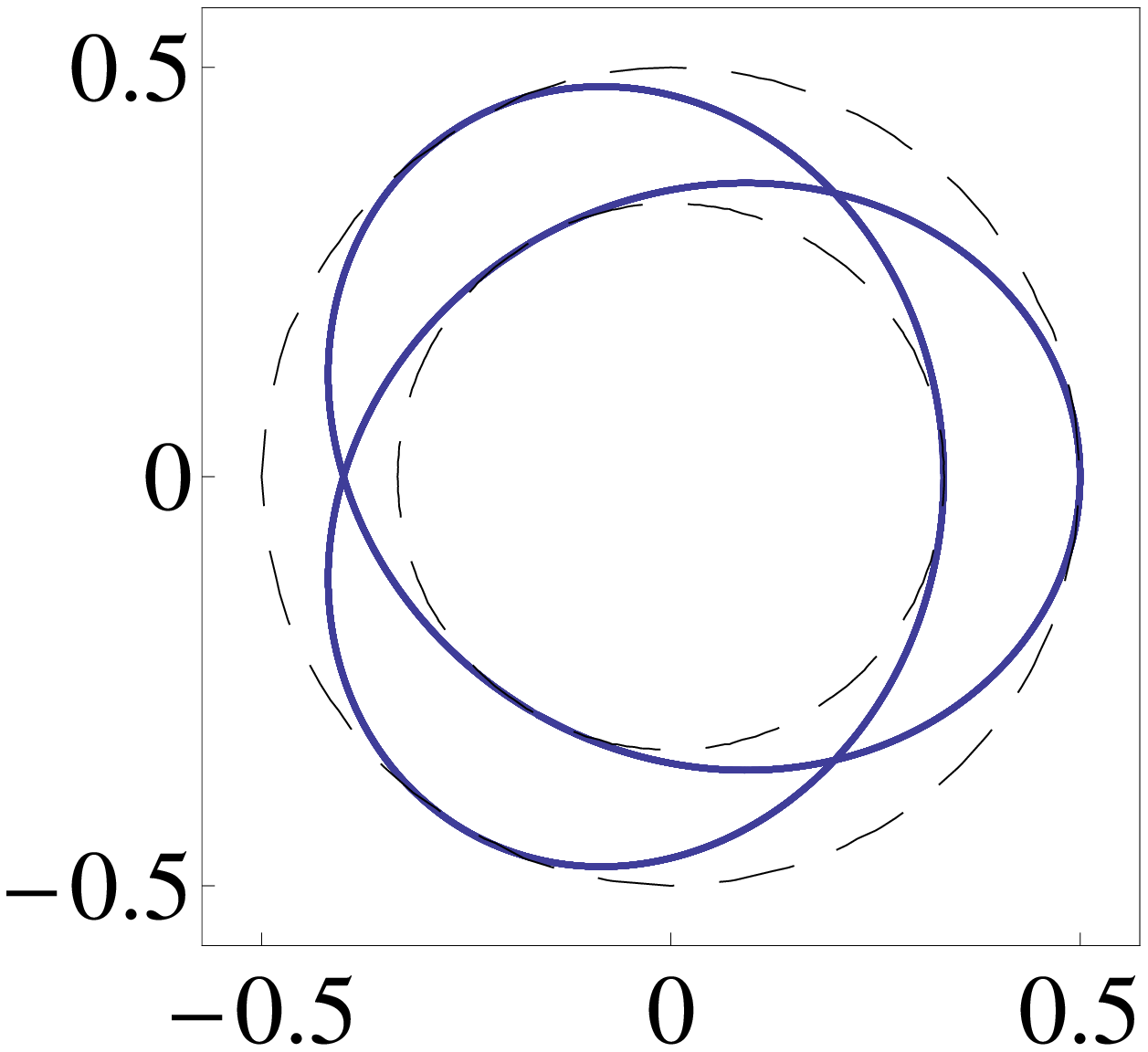}
\end{center}
\end{minipage}
\end{tabular}
\begin{flushleft}
\qquad \quad (b) $l_1=3/8,~l_2=-1/24$, \\
\qquad \qquad $(k,m; \bar k,\bar m)=(3,1; 3,-2)$
\end{flushleft}
\end{minipage}
\begin{minipage}[]{0.33\hsize}
\begin{tabular}{c}
\begin{minipage}[]{1\hsize}
\begin{center}
\includegraphics[width=\figsize,clip]{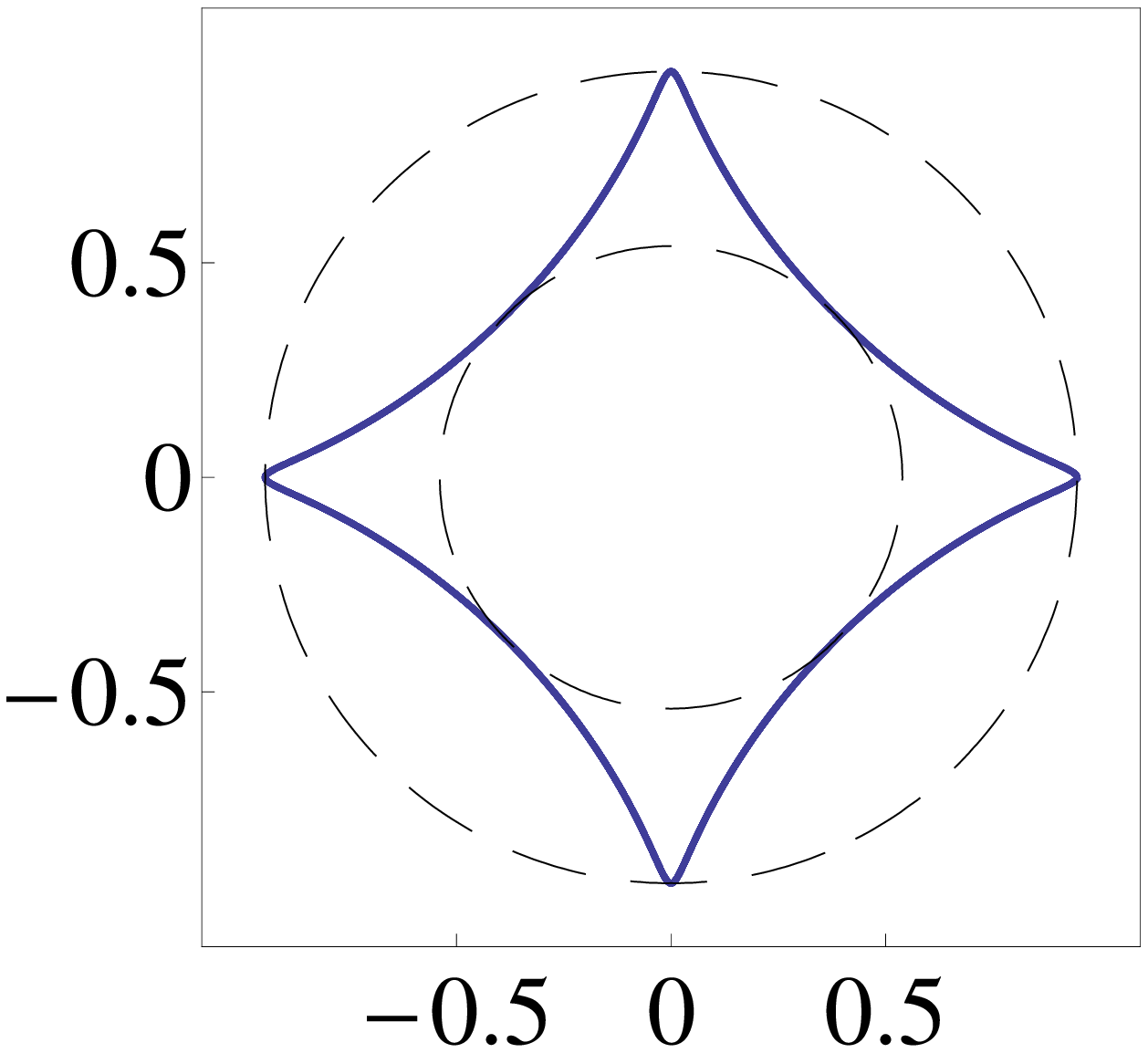}
\end{center}
\end{minipage}
\\
\begin{minipage}[]{1\hsize}
\begin{center}
\includegraphics[width=\figsize,clip]{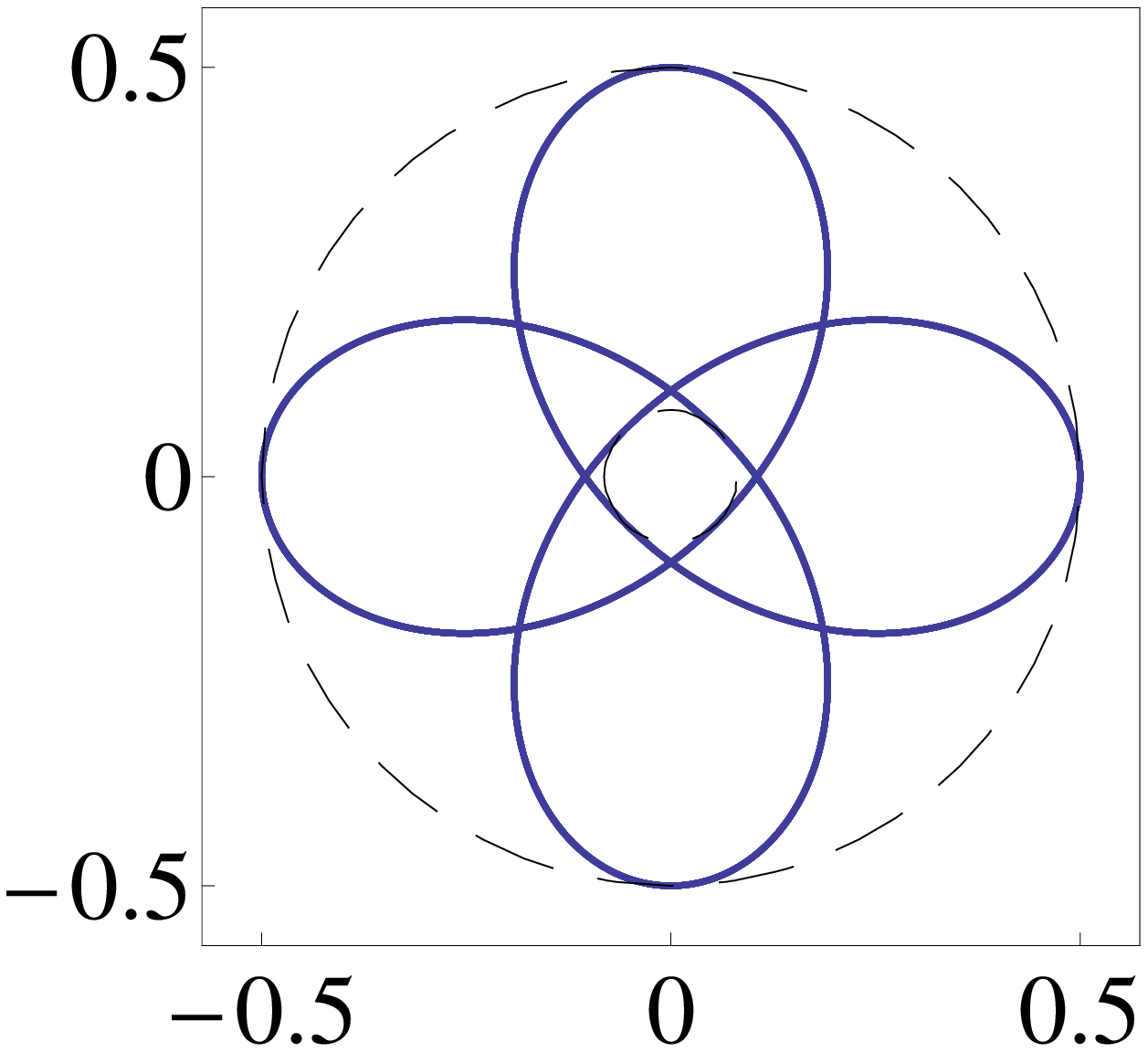}
\end{center}
\end{minipage}
\end{tabular}
\begin{flushleft}
\qquad \quad (c) $l_1=0.51,~l_2=-0.02$, \\
\qquad \qquad $(k,m; \bar k,\bar m)=(4,1; 4,-3)$
\end{flushleft}
\end{minipage}
\end{tabular}
\\
\bigskip
\bigskip
\begin{tabular}{ccc}
\begin{minipage}[]{0.33\hsize}
\begin{tabular}{c}
\begin{minipage}[]{1\hsize}
\begin{center}
\includegraphics[width=\figsize,clip]{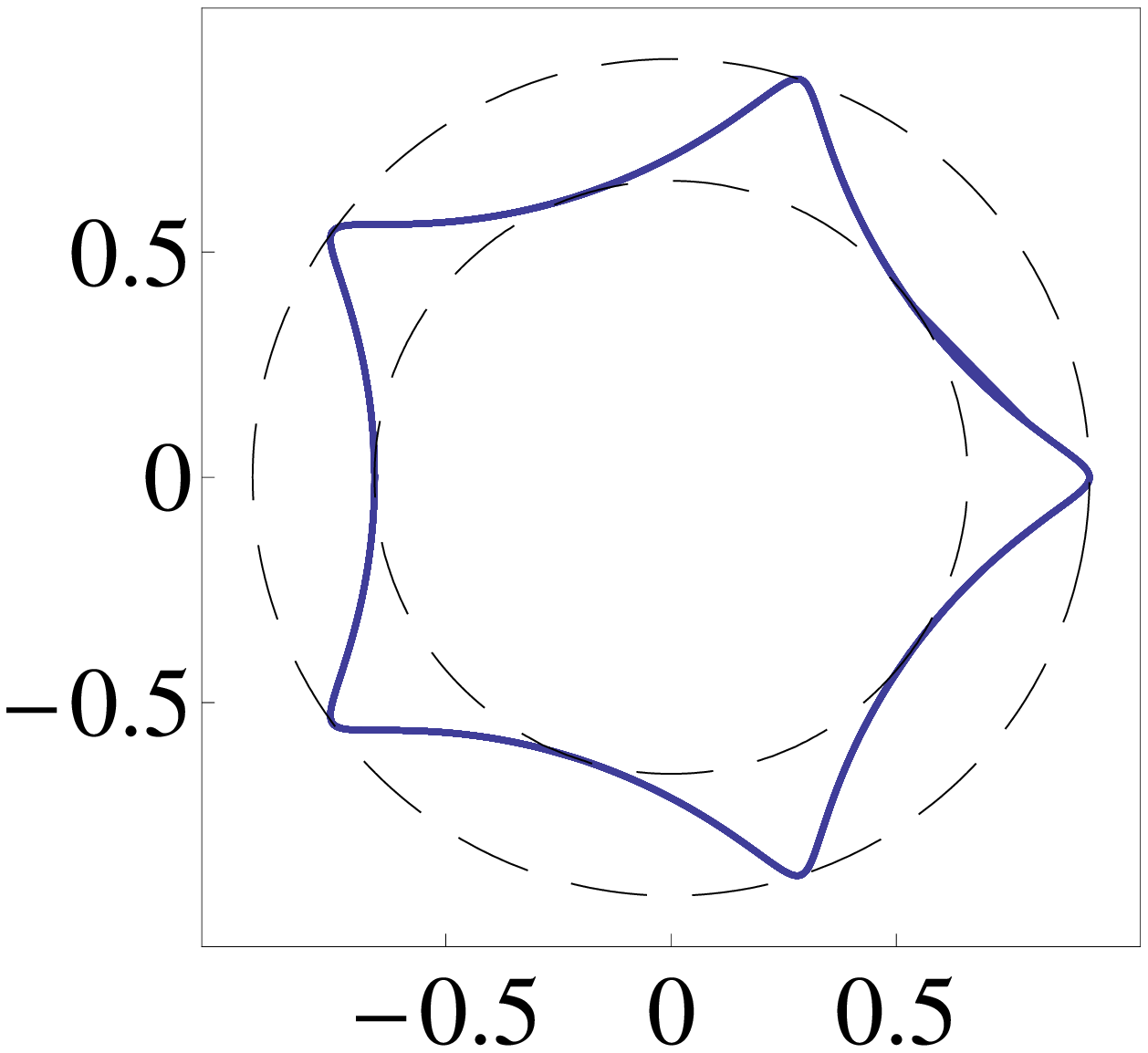}
\end{center}
\end{minipage}
\\
\begin{minipage}[]{1\hsize}
\begin{center}
\includegraphics[width=\figsize,clip]{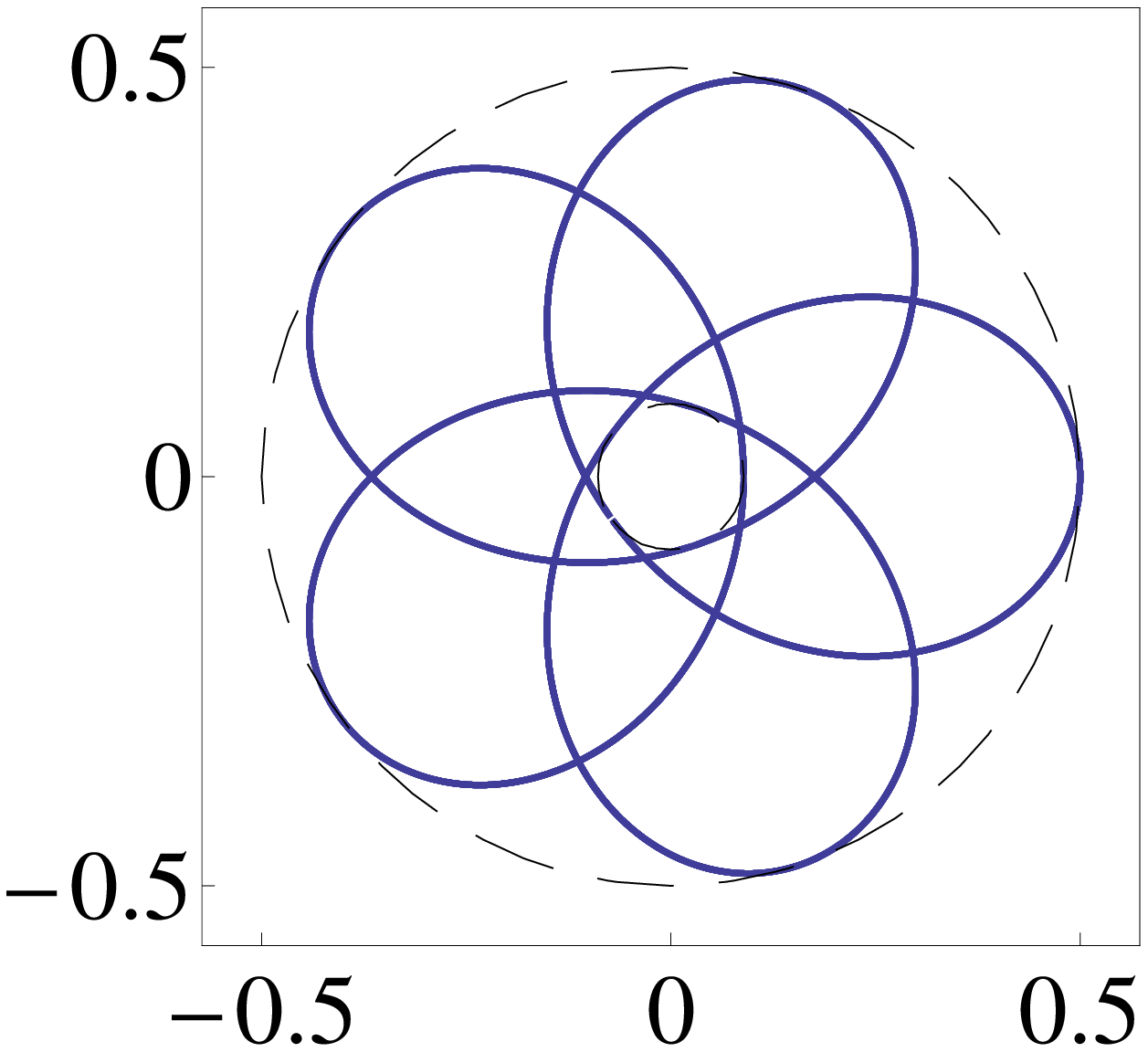}
\end{center}
\end{minipage}
\end{tabular}
\begin{flushleft}
\qquad \quad (d) $l_1=0.61,~l_2=-0.01$, \\
\qquad \qquad $(k,m; \bar k,\bar m)=(5,1; 5,-4)$
\end{flushleft}
\end{minipage}
\begin{minipage}[]{0.33\hsize}
\begin{tabular}{c}
\begin{minipage}[]{1\hsize}
\begin{center}
\includegraphics[width=\figsize,clip]{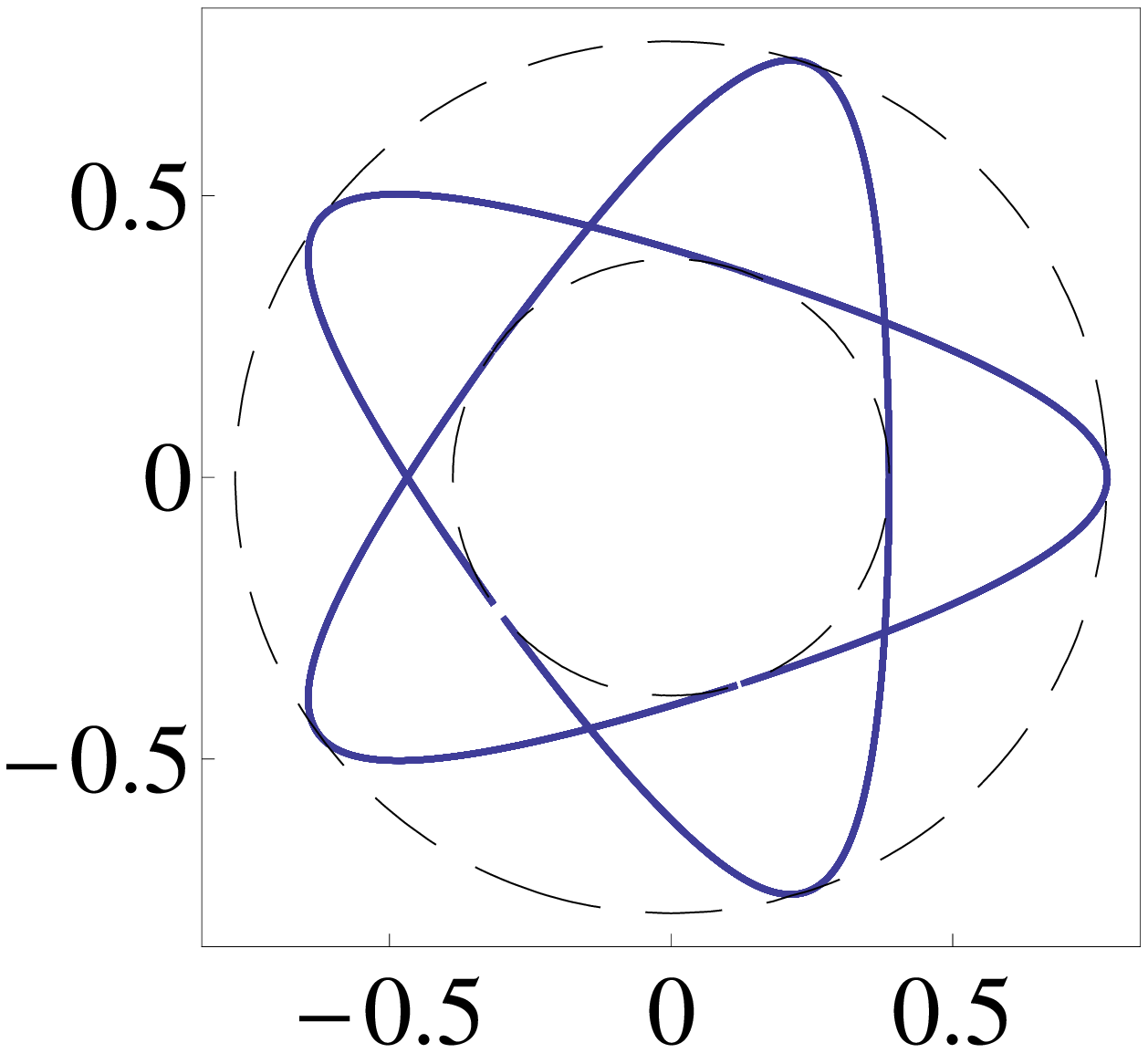}
\end{center}
\end{minipage}
\\
\begin{minipage}[]{1\hsize}
\begin{center}
\includegraphics[width=\figsize,clip]{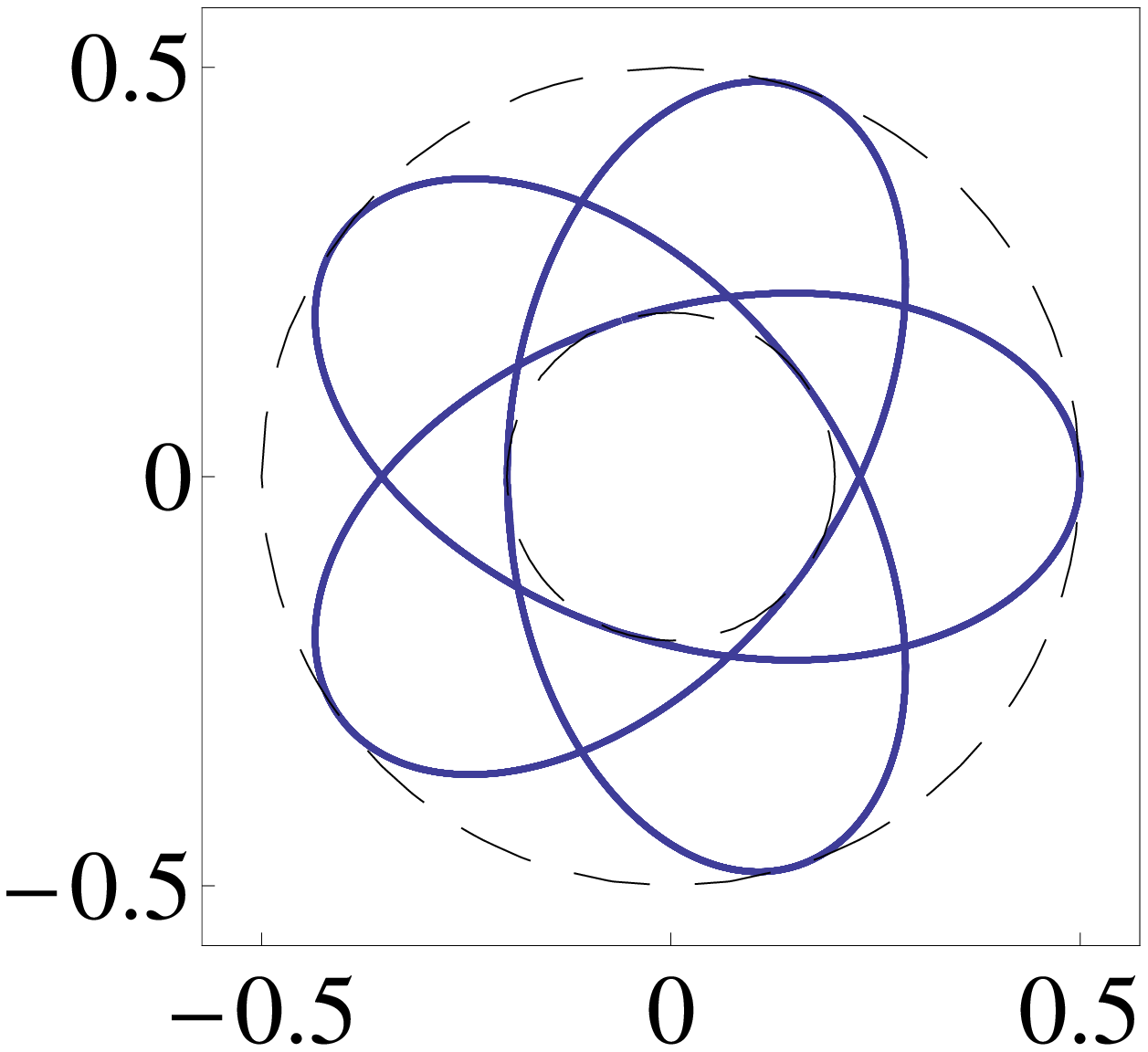}
\end{center}
\end{minipage}
\end{tabular}
\begin{flushleft}
\qquad \quad (e) $l_1=0.33,~l_2=-0.13$, \\
\qquad \qquad $(k,m; \bar k,\bar m)=(5,2; 5,-3)$
\end{flushleft}
\end{minipage}
\begin{minipage}[]{0.33\hsize}
\begin{tabular}{c}
\begin{minipage}[]{1\hsize}
\begin{center}
\includegraphics[width=\figsize,clip]{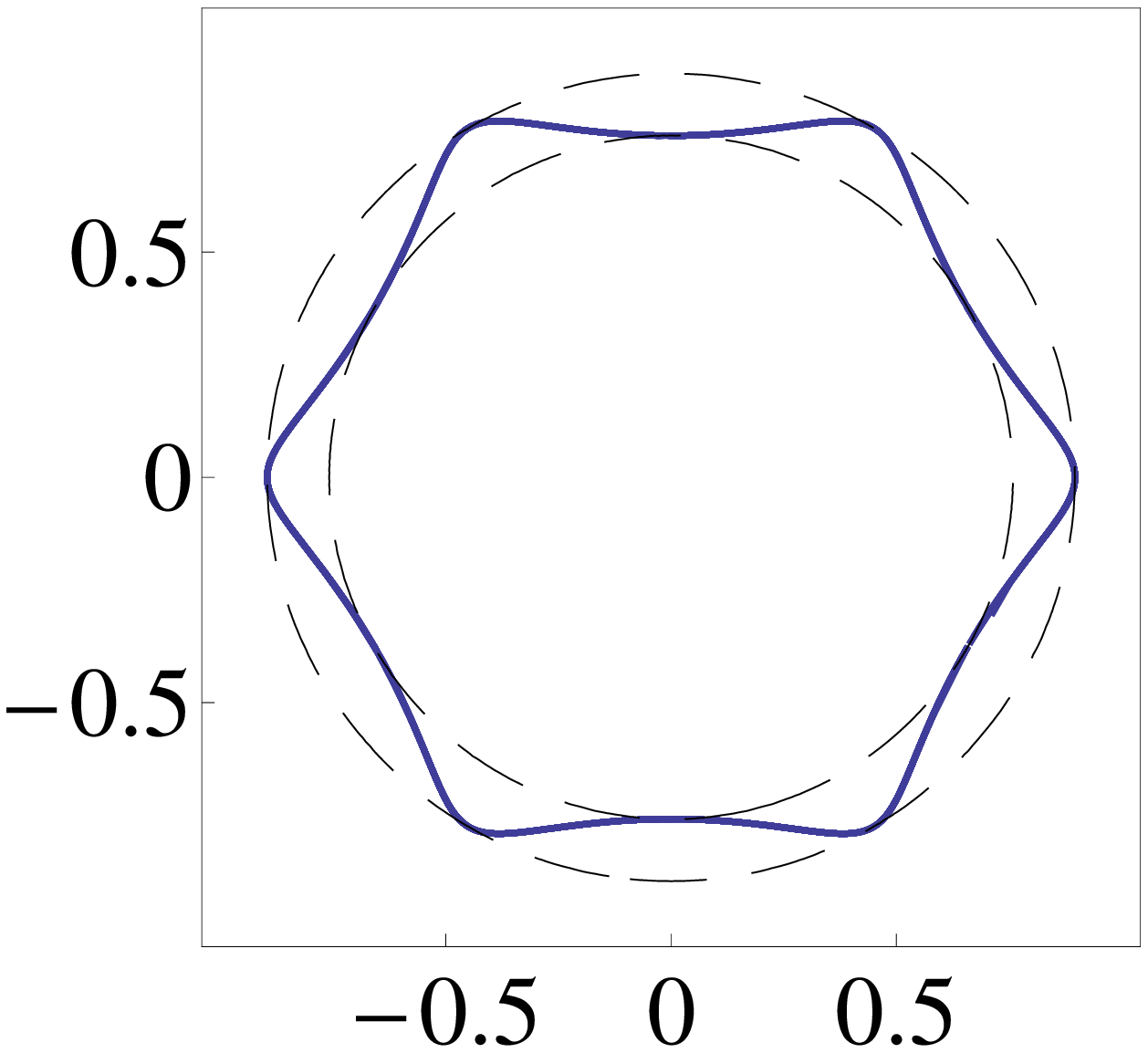}
\end{center}
\end{minipage}
\\
\begin{minipage}[]{1\hsize}
\begin{center}
\includegraphics[width=\figsize,clip]{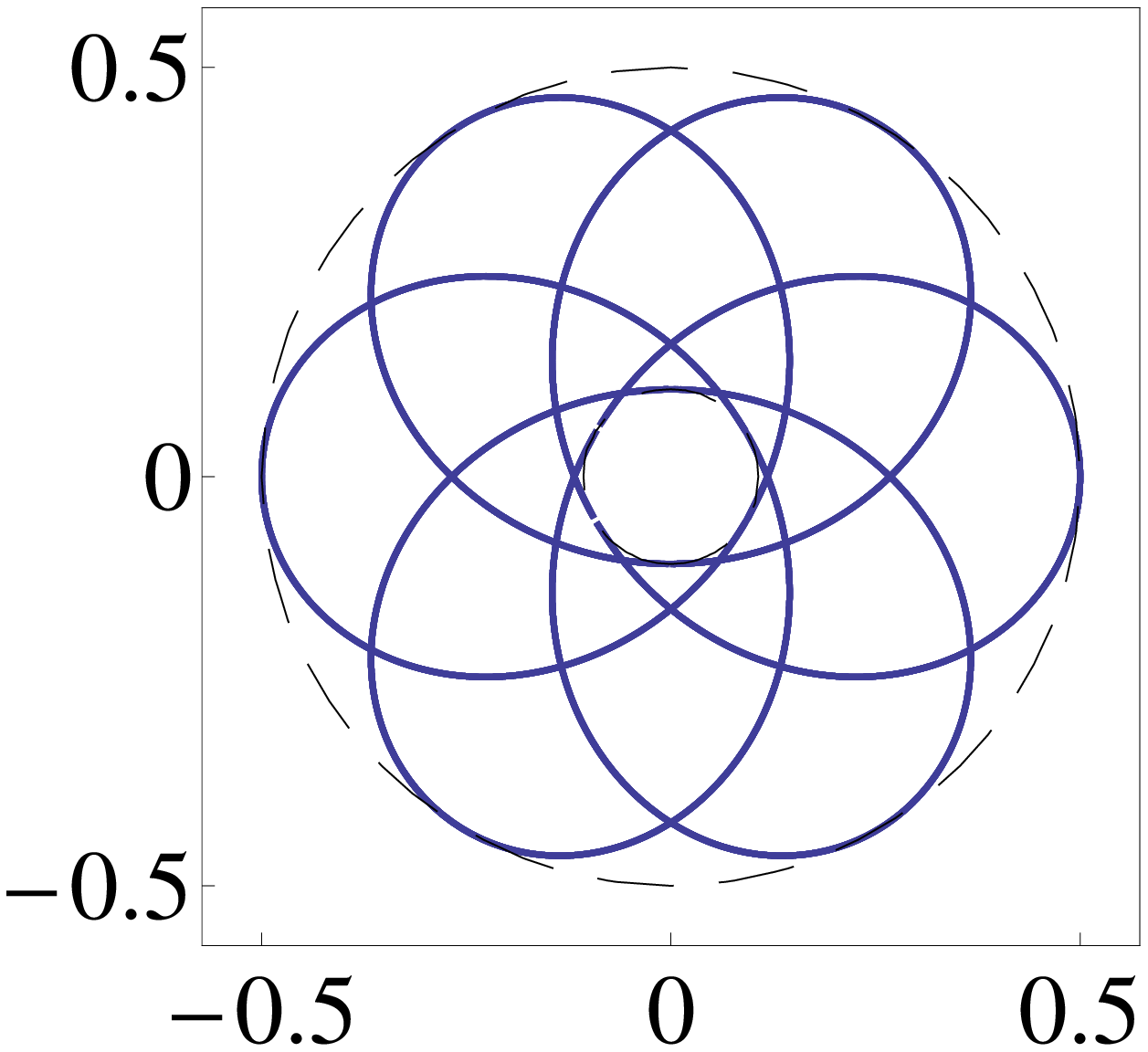}
\end{center}
\end{minipage}
\end{tabular}
\begin{flushleft}
\qquad \quad (f) $l_1=0.68,~l_2=-0.02$, \\
\qquad \qquad $(k,m; \bar k,\bar m)=(6,1; 6,-5)$
\end{flushleft}
\end{minipage}
\end{tabular}
\smallskip
\caption{The same as Fig.~\ref{fig:general_loop+} in the case $l_1>0$, $l_2<0$. We choose $\alpha=\beta=1$ for (a),(c) and (e), and $\alpha=1$, $\beta=1/2$ for (b),(d) and (f). The radius $\zeta_+$ is normalized to $\zeta_+=1/2$. }
\label{fig:general_loop-}
\end{figure}

\begin{center}
\begin{figure}[!h]
\begin{tabular}{ccc}
\begin{minipage}[]{0.33\hsize}
\begin{tabular}{c}
\begin{minipage}[]{1\hsize}
\begin{center}
\includegraphics[width=\figsize,clip]{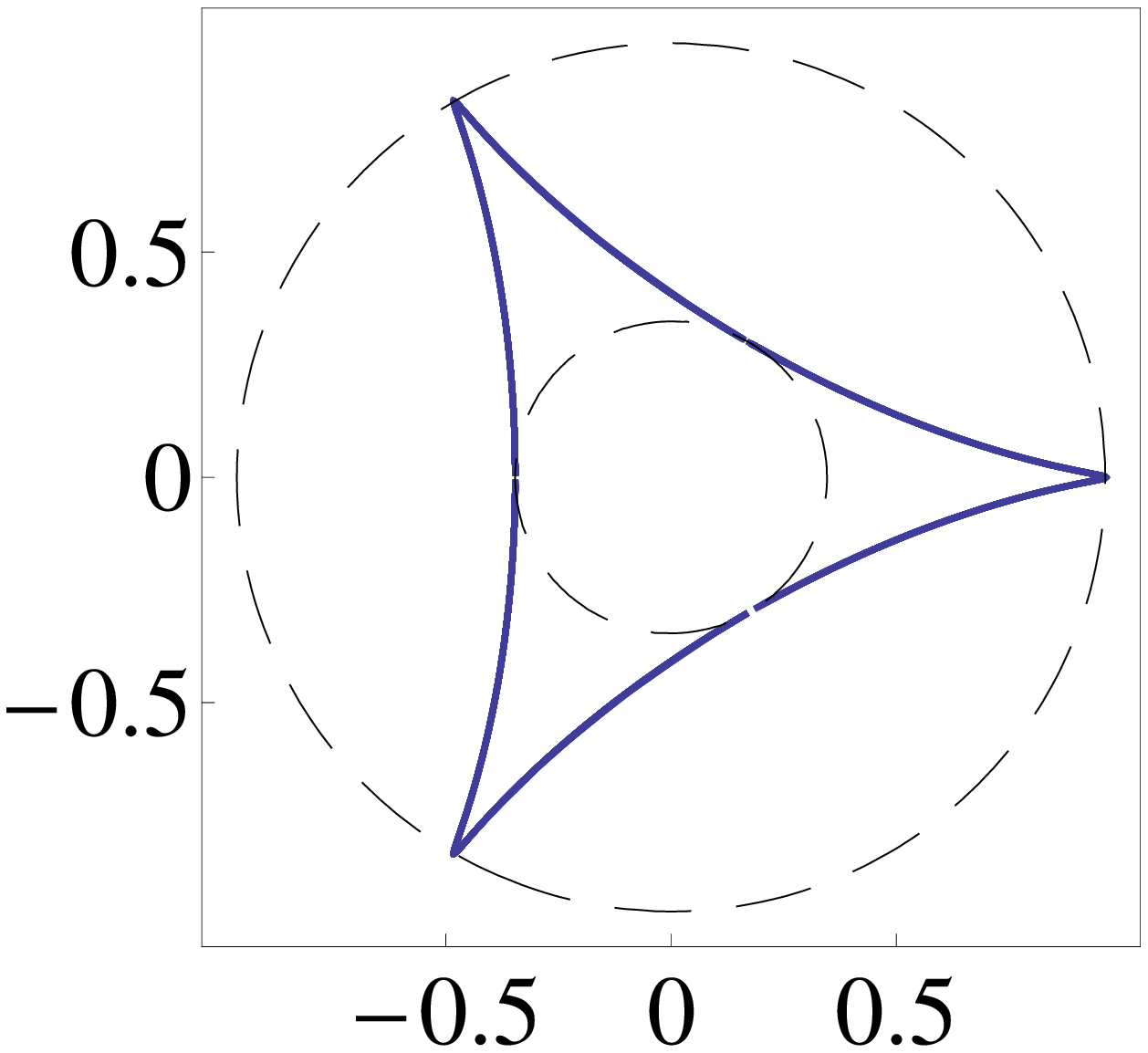}
\end{center}
\end{minipage}
\\
\begin{minipage}[]{1\hsize}
\begin{center}
\includegraphics[width=\figsize,clip]{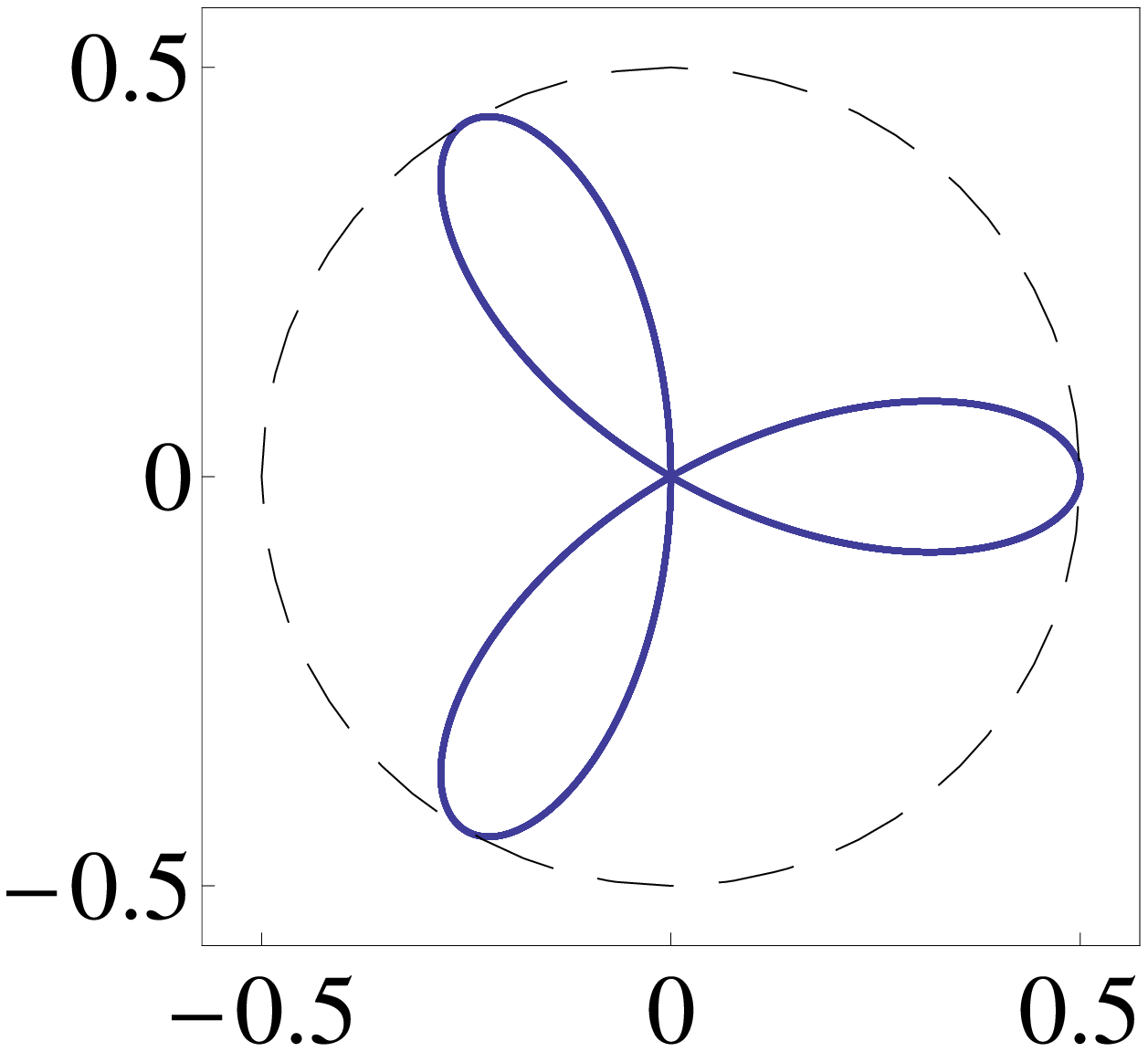}
\end{center}
\end{minipage}
\end{tabular}
\begin{flushleft}
\qquad \quad (a) $l_1=1/3,~l_2=0$, \\
\qquad \qquad $(k,m; \bar k, \bar m)=(3,1; 3,1)$\\
\hspace{89pt} $=(3,1; 3, -2)$
\end{flushleft}
\end{minipage}
\begin{minipage}[]{0.33\hsize}
\begin{tabular}{c}
\begin{minipage}[]{1\hsize}
\begin{center}
\includegraphics[width=\figsize,clip]{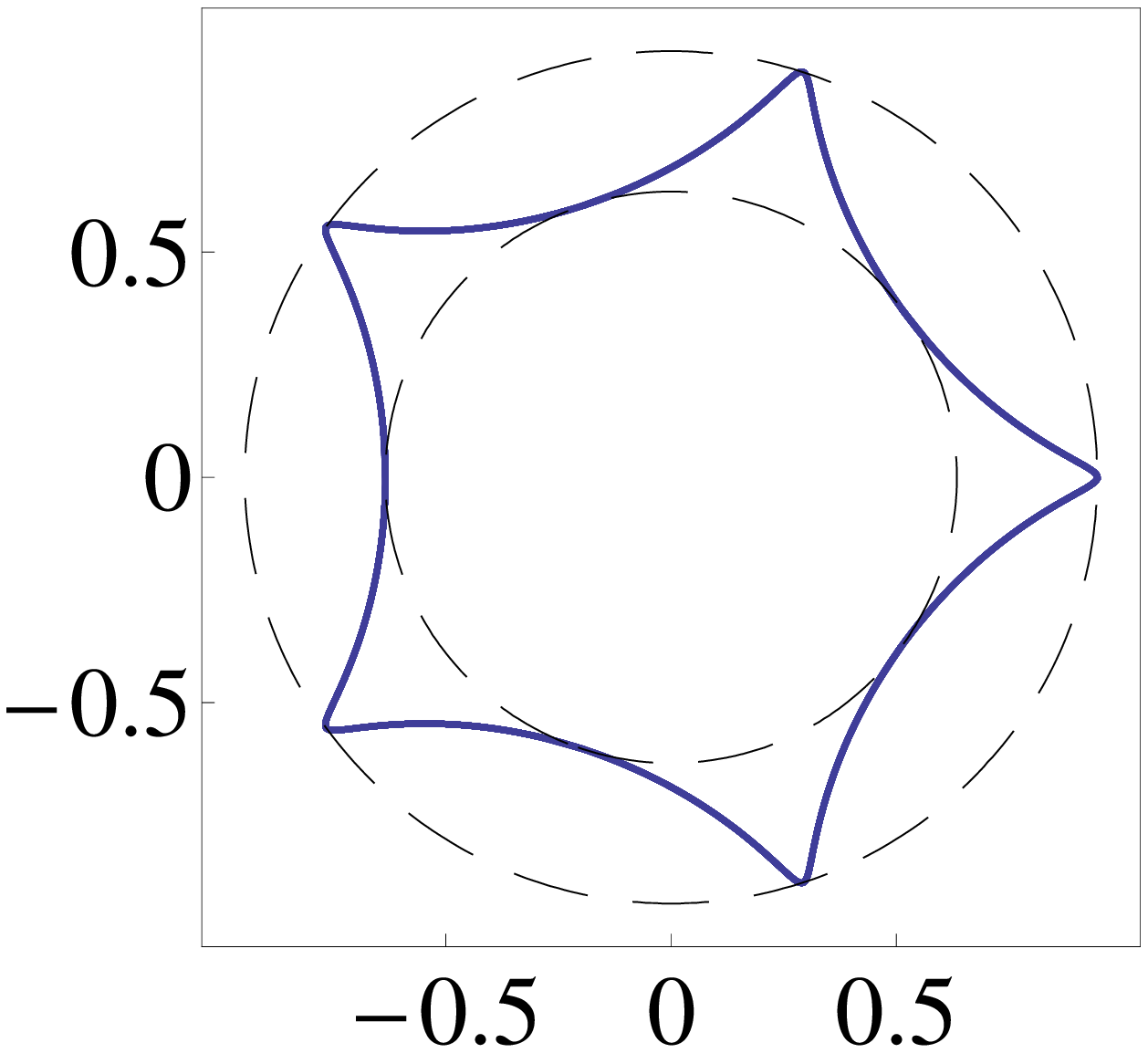}
\end{center}
\end{minipage}
\\
\begin{minipage}[]{1\hsize}
\begin{center}
\includegraphics[width=\figsize,clip]{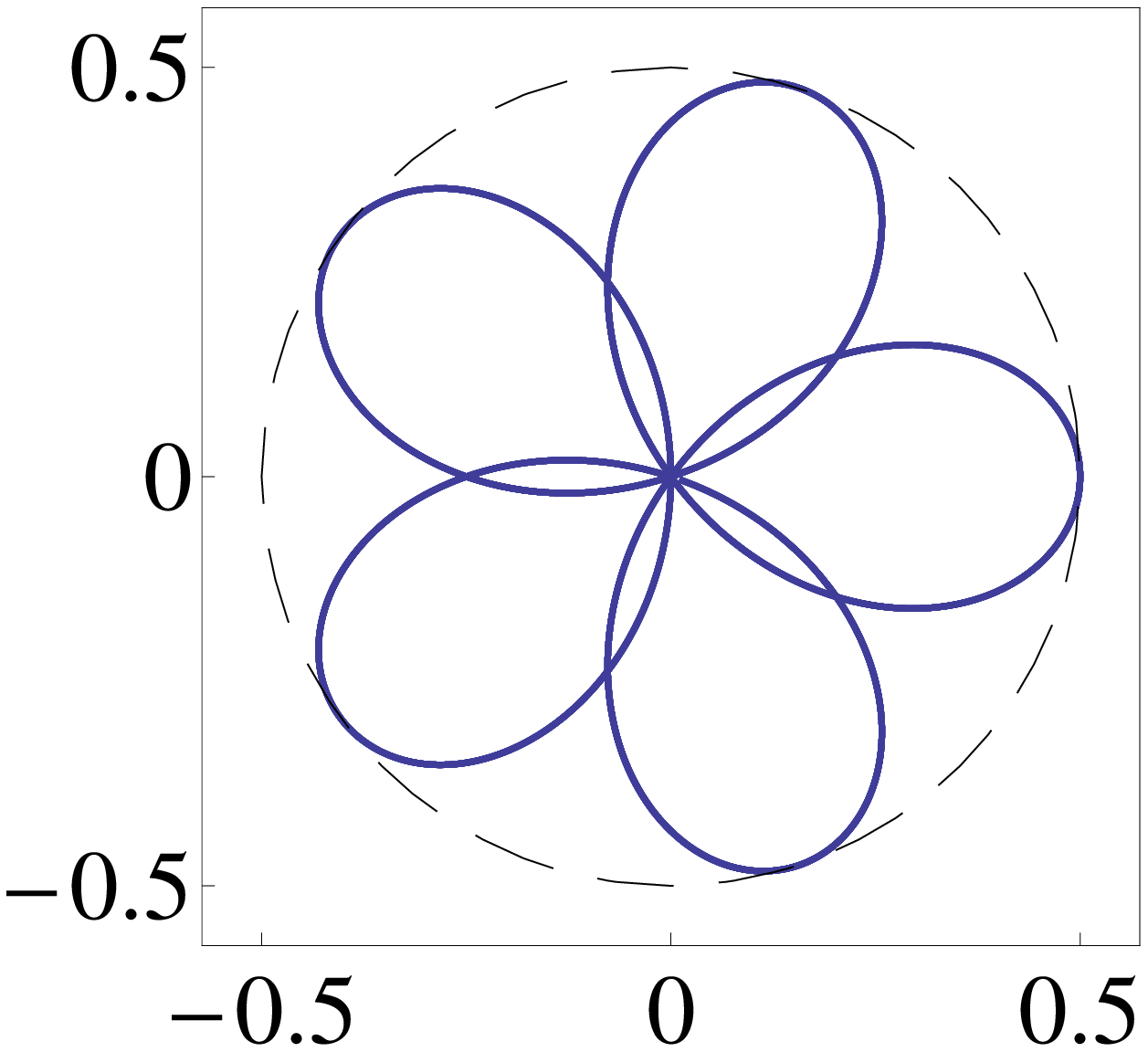}
\end{center}
\end{minipage}
\end{tabular}
\begin{flushleft}
\qquad \quad (b) $l_1=3/5,~l_2=0$, \\
\qquad \qquad $(k,m; \bar k, \bar m)=(5,1; 5,1)$\\
\hspace{89pt} $=(5,1; 5, -4)$
\end{flushleft}
\end{minipage}
\begin{minipage}[]{0.33\hsize}
\begin{tabular}{c}
\begin{minipage}[]{1\hsize}
\begin{center}
\includegraphics[width=\figsize,clip]{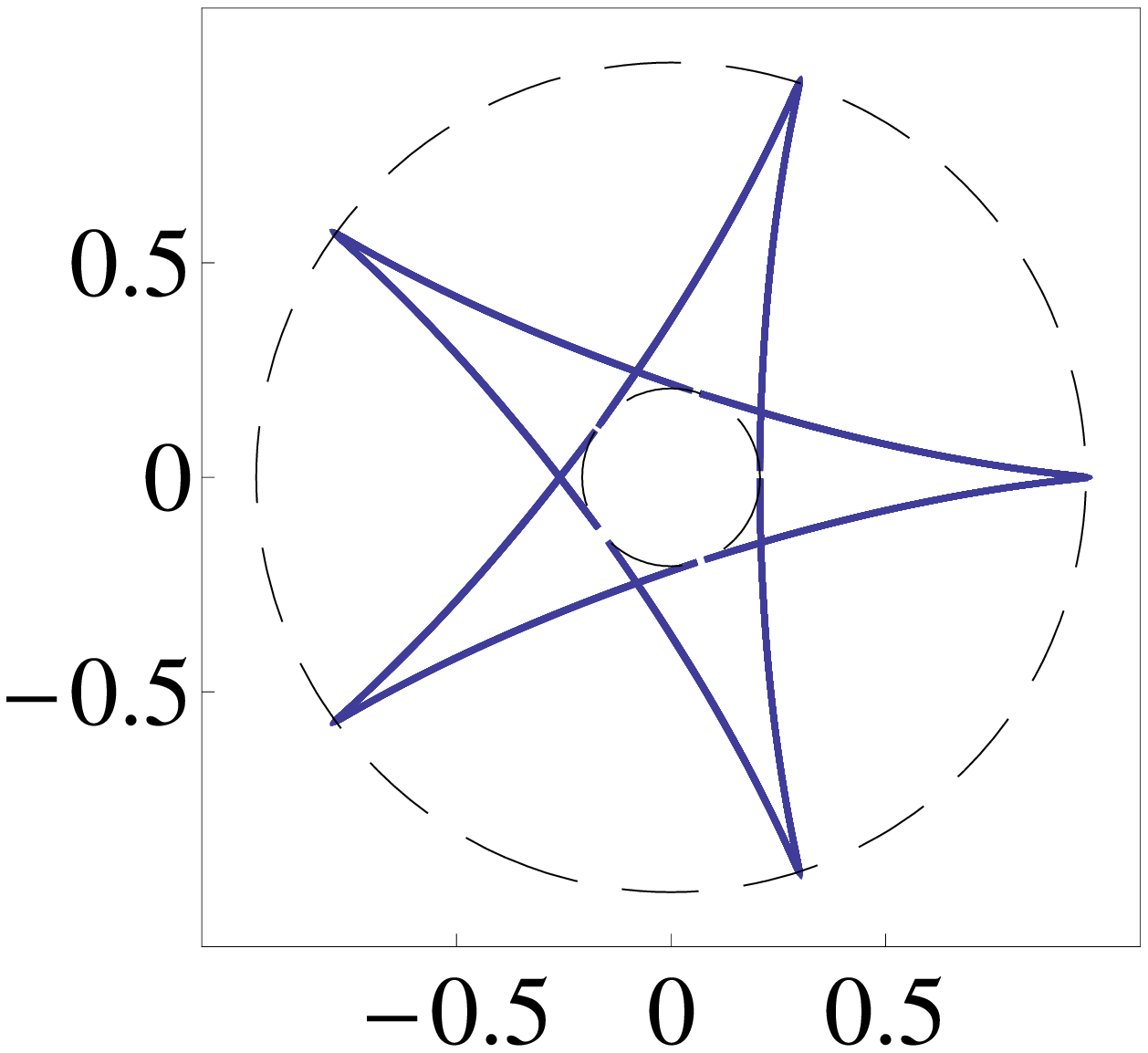}
\end{center}
\end{minipage}
\\
\begin{minipage}[]{1\hsize}
\begin{center}
\includegraphics[width=\figsize,clip]{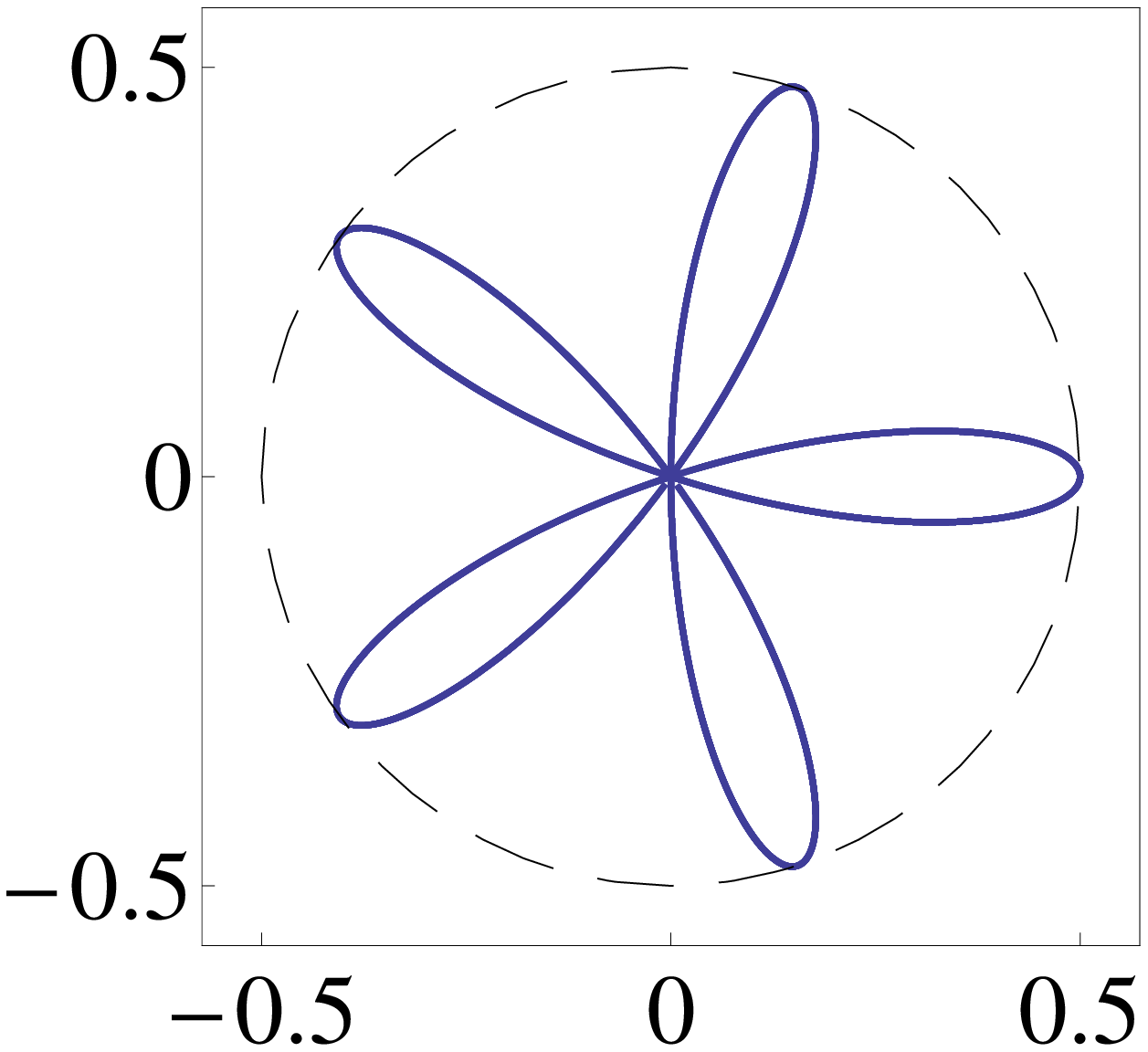}
\end{center}
\end{minipage}
\end{tabular}
\begin{flushleft}
\qquad \quad (c) $l_1=1/5, l_2=0$, \\
\qquad \qquad $(k,m; \bar k, \bar m)=(5,2; 5,2)$\\
\hspace{89pt} $=(5,2; 5, -3)$
\end{flushleft}
\end{minipage}
\end{tabular}
\caption{The same as Fig.~\ref{fig:general_loop+} in the case $l_1>0$, $l_2=0$. We choose $\alpha=1, \beta=1/2$. The radius $\zeta_+$ is normalized to $\zeta_+=1/2$.}
\bigskip
\label{fig:general_loop_zerol}
\end{figure}
\end{center}

{~}
\newpage
{~}
\section{Summary and Discussion}
In this paper, we have obtained general solutions for stationary rotating closed Nambu-Goto strings in five-dimensional flat spacetime. The worldsheet of the string admits a timelike Killing vector that is a linear combination of time translation Killing vector and two commutable rotation Killing vectors. The problem of finding solutions for the strings reduces to solving geodesic equation in the orbit space defined by projection with respect to the timelike Killing vector. We have found that separation of variables occurs in the Hamilton-Jacobi equation due to existence of residual two Killing vectors and a Killing tensor that are commutable each other in the orbit space. We have shown that a variety of stationary closed configurations of string can exist in the five-dimensional spacetime explicitly in contrast to the case in four-dimensional spacetime, where stationary closed strings are prohibited by the cusp formation.

A configuration of a stationary closed string is represented by a closed geodesic in the orbit space. In the five-dimensional flat spacetime spanned by the Cartesian coordinates $(T, X, Y, Z, W)$, there are two orthogonal planes, say $X$-$Y$ plane and $Z$-$W$ plane, on which the $X$-$Y$ rotation Killing vector commutes with the $Z$-$W$ rotation Killing vector. The stationary rotating closed string winds the center of the $X$-$Y$ plane while the string winds the center of the $Z$-$W$ plane. The winding string rotates in the $X$-$Y$ plane and rotates in the $Z$-$W$ plane, simultaneously. The stationary closed string configuration is achieved by the balance of the centrifugal repulsion of two independent rotations and the string tension.

We have discussed two special solutions: toroidal spiral strings and planar strings. The closed toroidal spiral string that lies on a two-dimensional torus, $\mathrm{S}^1$ on the $X$-$Y$ plane times $\mathrm{S}^1$ on the $Z$-$W$ plane~\cite{Igata:2009dr}. The toroidal spiral string has homogeneous worldsheet, that is, tangent to two linearly independent Killing vectors. The other special case is the planar strings that lies on a rotating two-dimensional plane. The planar strings are described by Lissajous figures on the plane, and have self-intersecting points in general, except for simple loops described by Lissajous ovals. Special closed loops which belong to both toroidal spirals and planar loops are simple loops which we called Hopf loops in Ref.~\cite{Igata:2009dr}. In the general cases, projection of snapshots of closed strings on the $X$-$Y$ and $Z$-$W$ planes are rounded polygons or rounded star polygons which are characterized by two sets of relatively primes. There exist lots of variation for stationary closed strings.

The string solutions obtained in the present paper is a stationary state for a test string. If we take gravitational interaction into account, the strings would emit gravitational waves~\cite{Ogawa:2008yx}. It would be expected that the closed strings in higher dimensions emit gravitational waves constantly, not burst like. The closed strings loose energy and angular momenta gradually by the emission. Therefore, it is an interesting problem to clarify the relation between shapes of closed strings and emission rates of gravitational waves, and evolutions of closed strings by gravitational wave emissions. The backreaction problem of gravitational wave emission to the strings is a challenging issue~\cite{Kodama:1994vb, Nakamura:2000eh, Nakamura:2000bw}. If the closed strings live long time in the universe, they would be a candidate of dark matter in the framework of higher-dimensional cosmology.

\acknowledgments
This work is supported by Grant-in-Aid for JSPS Fellows No.J111000492 (T.I.) and Grant-in-Aid for Scientific Research No.19540305 (H.I.).

\end{document}